\documentclass{ws-ijmpa}
\usepackage[compress]{cite} 
\usepackage{url}
\usepackage{hyperref}
\newcommand{\be}{\begin{eqnarray}}
\newcommand{\ee}{\end{eqnarray}}

\newcommand{\benum}{\begin{enumerate}}
\newcommand{\eenum}{\end{enumerate}}

\begin{document}

\markboth{C. Gale, S. Jeon, B. Schenke}
{HYDRODYNAMIC MODELING OF HEAVY-ION COLLISIONS}

%
\catchline{}{}{}{}{}
%

\title{HYDRODYNAMIC MODELING OF HEAVY-ION COLLISIONS}

\author{CHARLES GALE, SANGYONG JEON}
\address{Department of Physics, McGill University, 3600 University Street, Montreal, Quebec, H3A\,2T8, Canada}
\author{BJ\"ORN SCHENKE}
\address{Physics Department, Brookhaven National Laboratory, Upton, NY 11973, USA}



\maketitle


 \begin{abstract}
   We review progress in the hydrodynamic description of heavy-ion collisions,
   focusing on recent developments in modeling the fluctuating initial state and
   event-by-event viscous hydrodynamic simulations. 
   We discuss how hydrodynamics can be used to extract information on fundamental properties
   of quantum-chromo-dynamics from experimental data, and review successes and challenges of the hydrodynamic framework.
 \end{abstract}


\section{Introduction}
The large elliptic flow measured at the Relativistic Heavy-Ion Collider (RHIC) at Brookhaven National Laboratory and recently at the Large Hadron Collider (LHC) at CERN is one of the most striking observations in heavy-ion collision experiments.
This asymmetry of particle production in the transverse plane of the collision is interpreted as a sign of the system's hydrodynamic response to the initial geometry.
Ideal hydrodynamics, which includes no viscous effects,
was first used to describe heavy-ion collisions, and did a surprisingly good job in reproducing experimental data. Improved agreement with the data can be achieved using viscous hydrodynamic simulations with a very low shear viscosity to entropy density ratio. 
The applicability of hydrodynamics demands a short mean free path with respect to the system size.
Therefore it is concluded that the created quark-gluon plasma is strongly interacting and behaves like a nearly perfect fluid.

Recently interest has cascaded to higher harmonic flow, representing e.g. a triangular variation in the produced particle spectrum, which is non-zero in single events with lumpy initial energy density distributions. Hydrodynamic simulations together with models for the fluctuating initial state have been very successful in reproducing experimental data on all measured flow harmonics, their probability distributions and other quantities.

The aim in the application of hydrodynamics to heavy-ion collisions and comparison to experimental data is the extraction of properties of the created quark-gluon plasma and to learn about the initial state and its fluctuations. This provides information on fundamental properties of quantum chromo dynamic systems. Here, we review the tremendous progress that has been made over the last couple of years to achieve these goals.

\section{A short history}
We begin by reviewing the history of flow measurements at RHIC and LHC and the evolution of theory, in particular 
the hydrodynamic description of the bulk medium alongside experimental advances.

It was widely expected before the first RHIC results in 2000 that, because of the extremely high energy densities achieved in heavy-ion collisions
at $\sqrt{s}=200\,{\rm GeV}$ and the asymptotic freedom of QCD, the system would be one of weakly interacting partons. 
This system should behave like a gas and expand isotropically. However, the first results from RHIC demonstrated that this was not the case:
Particles emerging from heavy-ion collisions at RHIC showed a significant azimuthal anisotropy. 
This anisotropy, characterized by the second Fourier coefficient $v_2$ in an expansion of the azimuthal particle distribution, was 50\% larger
than that measured at SPS ($\sqrt{s}=17.3\,{\rm GeV}$), and agreed well with calculations using ideal hydrodynamics, initialized to
reproduce the measured multiplicity and transverse momentum spectra \cite{Kolb:2003dz,Huovinen:2003fa,Hirano:2002ds}.
The success of ideal hydrodynamics implied that the system was strongly interacting and very different from early expectations.

Both measurements and theoretical calculations became more sophisticated. Detailed measurements of $v_2$ for different particle species followed,
showing a clear mass-splitting, predicted by hydrodynamics and due to the emergence of all particles from a single velocity field.

The relevance of viscous corrections for the hydrodynamic description of heavy-ion collisions was never completely ignored, but it was not until
the ratio of viscosity over entropy density in strongly-coupled systems was calculated using AdS/CFT techniques \cite{Kovtun:2004de} in 2004, in a regime where standard kinetic theory was known to break down, that the study of viscous effects became an urgent matter.
The AdS/CFT calculations predicted that the viscosity to entropy density ratio of an $N=4$ super Yang-Mills quantum system in the strong coupling limit is $\eta/s=1/4\pi$, a value that can be rationalized by the argument that excitations cannot be localized with a precision smaller than their thermal wavelength \cite{Danielewicz:1984ww}, but which had not been reliably calculated previously.

From then on viscous hydrodynamic calculations and comparison to experimental data emerged as the potentially best way to determine the transport properties of the matter created in heavy-ion collisions. Earlier perturbative QCD calculations predicted a rather large value of $\eta/s$ (compared to $1/4\pi$) with, however, significant error bands, mainly resulting from uncertainties in the relevant scales. \cite{Arnold:2003zc} Non-perturbative lattice QCD calculations existed \cite{Nakamura:2004sy,Meyer:2007ic,Meyer:2009jp}, but because transport properties are dynamical quantities, some doubts remain concerning their reliability.
The quest to determine the effective $\eta/s$ with precision began and recent studies \cite{Luzum:2012wu} find it in the range $0.07 \leq \eta/s \leq 0.43$. This result is for an average value that corresponds to the viscosity at the typical time when flow is built up. \cite{Luzum:2012wu,Song:2012ua}
Some effort has been directed at developing a framework for precision determination of different parameters, like this effective shear viscosity to entropy density ratio, in a detailed multi-parameter $\chi^2$ fit to multiple experimental quantities \cite{Soltz:2012rk}. There are, however, missing ingredients like the fluctuations described below such that it is probably too early to attempt precision fits to experimental data.
Future goals along these lines naturally include the determination of the temperature dependence of $\eta/s$ and other parameters, such as bulk viscosity.
We will discuss these endeavors in more detail in the following sections.

\begin{figure}[!t]
\begin{center}
 \includegraphics[width=0.93\textwidth]{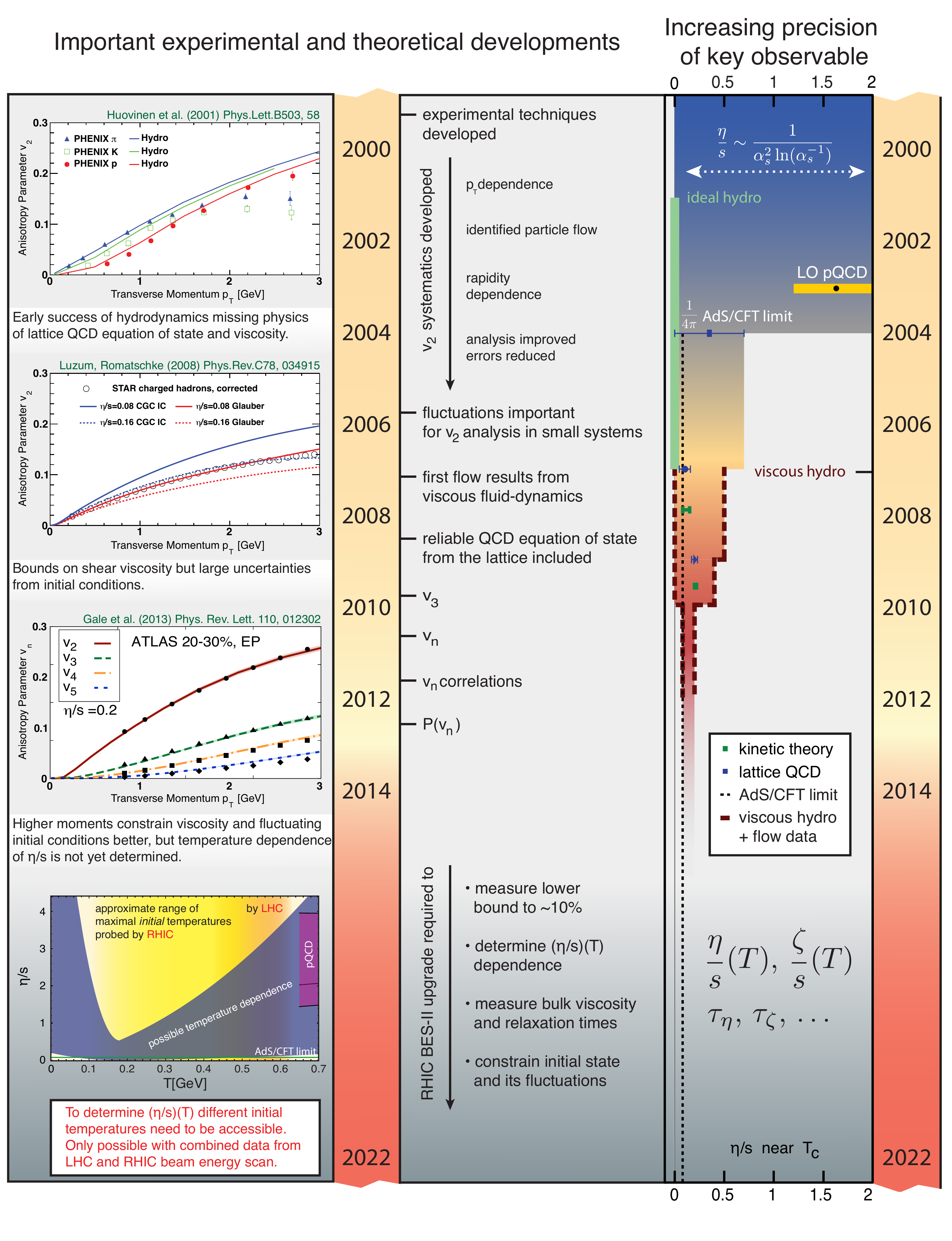}
 \caption{Time line of important experimental and theoretical developments leading towards increasingly precise understanding of flow, 
   transport properties of the quark-gluon plasma, and the initial state and its fluctuations. On the left are three key figures \cite{Huovinen:2001cy,Luzum:2008cw,Schenke:2011bn} depicting the progress of hydrodynamic calculations and their success in describing experimental data, followed by a sketch of the uncertainty in a temperature dependent $(\eta/s)(T)$. On the right, the increasing precision
   in one key observable, the shear viscosity to entropy density ratio $\eta/s$ near its minimal value, is illustrated (see text for details). Figure adapted from the ``Hot \& Dense QCD White Paper'', solicited by the NSAC subcommittee on Nuclear Physics funding in the US.
  \label{fig:timeline} }
\end{center}
\end{figure}
While theoretical work was focused on the study of viscous effects, experiments discovered the importance of fluctuations in the initial geometric configuration while studying Cu+Cu collisions.
Copper ions were initially used in 2005 to study whether interesting physics effects found in Au+Au collisions would turn off. Interestingly,
$v_2$ was found to be quite large, even in central Cu+Cu collisions \cite{Alver:2006wh}, where $v_2$ should be very close to zero if the initial energy density 
could be described as a convolution of two smooth nucleon distributions at impact parameter $b\approx 0\,{\rm fm}$. 
The concept of the ``participant eccentricity'', where the shape of the overlap region was not calculated relative to the classical impact parameter, but relative to an axis determined by the fluctuating participants, allowed for an explanation of the observed large $v_2$.

Fluctuations are also the reason why odd flow harmonics (such as $v_3$, $v_5$, etc.) are not zero. It was not until 2010 that real interest
in the study of higher flow harmonics was sparked \cite{Alver:2010gr,Alver:2010dn}, even though their potential was pointed out earlier \cite{Mishra:2007tw}.
First results on higher-order harmonic flow by the RHIC and LHC experiments appeared just before and at the Quark Matter 2011 conference.
$v_1$ through $v_6$ were shown to be sizable, each having its own amplitude and event plane angle. Odd flow harmonics were found to have a weak centrality dependence, characteristic of initial state geometric fluctuations. Furthermore, the $v_n$ become weaker with increasing $n$, as expected from the presence of viscosity, which more efficiently damps out higher order (smaller wavelength) fluctuations.

The current state-of-the-art viscous hydrodynamic calculations with the most advanced models for the fluctuating initial state show remarkable agreement with flow measurements from both RHIC and LHC \cite{Gale:2012rq}. Even the event-by-event distributions of $v_2$-$v_4$ as measured by the ATLAS collaboration \cite{Jia:2012ve} are reproduced by these calculations.

The historical developments outlined in this section are summarized in a time line in Fig. \ref{fig:timeline}, presenting key figures and the increasing precision in the determination of an effective shear viscosity to entropy density ratio of hot and dense nuclear matter and an outlook to future goals, including the determination of the temperature dependence of shear and bulk viscosity as well as more details on other transport parameters.
The center column lists important developments as a function of time, including an outlook towards potential achievements in the next decade.
In the left column, we first highlight the early success of hydrodynamics in describing results on transverse momentum dependent elliptic flow from RHIC \cite{Huovinen:2001cy}. The simulation described the data very well despite the fact that important ingredients such as a QCD equation of state and viscous effects were neglected. This early success was however reason to pursue further developments within the hydrodynamic framework, leading to more and more sophisticated modeling. One such important step is represented by the second figure, showing the comparison of first viscous hydrodynamic calculations with different initial conditions and different shear viscosity to entropy density ratios to experimental data. The third figure shows the comparison of experimental data on $v_2(p_T)$, $v_3(p_T)$, $v_4(p_T)$, and $v_5(p_T)$ from the ATLAS collaboration \cite{ATLAS:2012at} to results from an event-by-event viscous hydrodynamic simulation including a QCD based model for the initial state fluctuations and using a constant $\eta/s=0.2$, showing striking agreement \cite{Gale:2012rq}. The final figure is a sketch of the uncertainty in the temperature dependent value of $\eta/s$, indicating that a combined analysis of LHC data and RHIC data at different energies using state-of-the-art hydrodynamic simulations could help determine this temperature dependence as well as that of other transport parameters.
On the right, the increasing precision in one key observable, the shear viscosity to entropy density ratio $\eta/s$ near its minimal value, is illustrated. Shown results are from perturbative QCD calculations at leading logarithmic order \cite{Hosoya:1983xm,vonOertzen:1990ad,Thoma:1991em,Arnold:2000dr} (shown formula), complete leading order \cite{Arnold:2003zc} (band from varying the scale by 20\%), Anti-deSitter gravity/Conformal Field Theory (AdS/CFT) correspondence \cite{Kovtun:2004de}, lattice QCD - pure glue at $\sim 1.6\,T_c$, $1.24\,T_c$, and $1.58\,T_c$, respectively  \cite{Nakamura:2004sy,Meyer:2007ic,Meyer:2009jp}, ideal hydrodynamics \cite{Kolb:2000sd,Kolb:2000fha}, perturbative QCD/kinetic theory \cite{Xu:2007jv,Chen:2009sm}, and viscous hydrodynamics constrained by flow measurements \cite{Romatschke:2007mq,Luzum:2008cw,Song:2008hj,Song:2010mg}. We indicate that going forward, efforts should focus on the determination of temperature dependent quantities since more precise values of an effective $\eta/s$ depend on the collision energy and other parameters of the collision.


Despite the great success of relativistic viscous hydrodynamics in describing experimental data from heavy-ion experiments, several puzzles remain.
It is for example still not understood how the system can thermalize quickly enough to allow for the early applicability of hydrodynamics (which is needed to describe the data), or whether this early thermalization is even needed at all. We will discuss this and other open questions and issues in the remainder of this review.

\section{Theoretical framework}
The current standard for the viscous hydrodynamic description of relativistic heavy-ion collisions has been established
with the derivation of hydrodynamic equations including all terms up to second order in gradients for a conformal fluid \cite{Baier:2007ix}. 
Additional terms for non-conformal fluids with non-zero bulk viscosity have been derived subsequently \cite{Betz:2009zz}.

In the ideal (non-viscous) case, the evolution of the system created in relativistic heavy-ion collisions
is described by the following 5 conservation equations
\begin{eqnarray}
& \partial_\mu T_{\rm id}^{\mu\nu} = 0\,, ~~\partial_\mu J_B^\mu = 0\,, \label{conservationEqns}
\end{eqnarray}
where $T_{\rm id}^{\mu\nu}$ is the energy-momentum tensor and $J_B^\mu$ is the net baryon
current. These are usually re-expressed
using the time-like flow four-vector $u^\mu$ as
\begin{eqnarray}\label{Tideal}
& T_{\rm id}^{\mu\nu} = (\varepsilon + \mathcal{P})u^\mu u^\nu - \mathcal{P} g^{\mu\nu}\,, ~~ J_{B}^{\mu} = \rho_{B} u^\mu \,,
\end{eqnarray}
where $\varepsilon$ is the energy density, $\mathcal{P}$ is the pressure, $\rho_B$
is the baryon density and 
$g^{\mu\nu} = \hbox{diag}(1, -1, -1, -1)$ is the metric tensor. 
The equations are then closed by adding the equilibrium 
equation of state
\begin{eqnarray}\label{eos}
\mathcal{P} = \mathcal{P}(\varepsilon, \rho_B)
\end{eqnarray}
as a local constraint on the variables.

In the first-order, or Navier-Stokes formalism for viscous hydrodynamics,
the stress-energy tensor is decomposed into
$
T_{\rm 1st}^{\mu\nu} = T^{\mu\nu}_{\rm id} + S^{\mu\nu}\,,
$
where
$T_{\rm id}^{\mu\nu}$ is given in Eq.\,(\ref{Tideal}) and the viscous part of the stress energy tensor is
\be
S^{\mu\nu} = \eta 
\left(
\nabla^\mu u^\nu + \nabla^\nu u^\mu - 
{2\over 3}\Delta^{\mu\nu}\nabla_\alpha u^\alpha
\right)\,,
\ee
where $\Delta^{\mu\nu} = g^{\mu\nu} - u^\mu u^\nu$ is the local 3-metric 
and $\nabla^\mu = \Delta^{\mu\nu}\partial_\nu$ is the local spatial derivative.
Note that $S^{\mu\nu}$ is transverse with respect to the flow velocity since
$\Delta^{\mu\nu}u_\nu = 0$ and $u^\nu u_\nu = 1$. 
Hence, $u^\mu$ is a (timelike)
eigenvector of the whole stress-energy tensor with the same eigenvalue
$\epsilon$. $\eta$ is the shear viscosity of the medium.

The Navier-Stokes form is conceptually simple but introduces unphysical 
super-luminal signals 
that lead to numerical instabilities. 

The second-order Israel-Stewart formalism \cite{Israel:1976tn,Stewart:1977,Israel:1979wp}
avoids this super-luminal propagation, as do more recent approaches
\cite{Muronga:2001zk,Denicol:2012es}. 
In the Israel-Stewart formalism for a conformal fluid \cite{Baier:2007ix}, the stress-energy tensor is decomposed as
$
{\cal T}^{\mu\nu} = T^{\mu\nu}_{\rm id} + \pi^{\mu\nu}\,.
$
The evolution equations are
\begin{eqnarray}
~~~~~~~~~~~\partial_\mu {\cal T}^{\mu\nu} &=& 0\,,\\
\Delta^{\mu}_{\alpha}\Delta^{\nu}_{\beta}
{u^\sigma\partial_\sigma} \pi^{\alpha\beta}
&=&
-{1\over \tau_\pi}
\left( \pi^{\mu\nu} - S^{\mu\nu} \right) - {4\over 3}\pi^{\mu\nu}(\partial_\alpha u^\alpha)\,,
\end{eqnarray}
where for brevity we neglected to show heat-flow, vorticity and terms that turn out to be numerically irrelevant. The role of vorticity in heavy-ion collisions when including fluctuations has also recently been studied \cite{Florchinger:2011qf}.

We note that the Israel-Stewart formalism is based on two main choices: The 14-moment approximation to truncate the
single particle distribution function, and the choice of the equations of motion for the dissipative currents \cite{Denicol:2012es}, which is not unique.
We will come back to this point when discussing recent theoretical developments in Section \ref{sec:beyond2nd}.

Simulations of bulk dynamics in heavy-ion collisions using the Israel-Stewart formalism have been performed in 2+1 dimensions \cite{Romatschke:2007jx,Romatschke:2007mq,Luzum:2008cw,Heinz:2005bw,Song:2007fn,Song:2008si}, assuming boost-invariance in the longitudinal direction. This is a reasonable assumption 
around mid-rapidity for high enough energy collisions, particularly at LHC energies.
2+1 dimensional simulations were also performed using the equivalent
\"Ottinger-Grmela \cite{Grmela:1997zz} formalism \cite{Dusling:2007gi}. More recently, 3+1 dimensional viscous calculations that are required away from mid-rapidity and at lower collision energies have also become available \cite{Schenke:2010rr,Schenke:2011tv,Bozek:2011ua}.

Having established the standard theoretical framework, 
in the following we will discuss recent developments in the field of relativistic viscous hydrodynamics and its application to heavy-ion collisions.


\section{Equation of state}
Typically, the equation of state (\ref{eos}) used in hydrodynamic simulations of heavy-ion collisions is determined from lattice QCD calculations combined with a hadron gas model. A detailed comparison of different lattice equations of state has been performed \cite{Huovinen:2009yb} and it was found that
the different equations of state \cite{Laine:2006cp,Chojnacki:2007jc,Chojnacki:2007rq,Bazavov:2009zn,Song:2008si}, which have different 
results for the trace anomaly and the speed of sound, lead to different evolution of the momentum anisotropy when used in an ideal hydrodynamic simulation.
However, the difference in the final spectra and elliptic flow results turns out to be negligible. This and uncertainties in the initial state make a precision determination of the equation of state from the comparison of hydrodynamic calculations to hadronic experimental data impossible. 
On the other hand, using the latest lattice equations of state within hydrodynamic calculations can potentially allow for the reliable extraction of medium properties such as transport coefficients from hadronic observables.
It is particularly encouraging that lattice QCD calculations from the HotQCD and Wuppertal-Budapest collaborations now agree on the value of the (pseudo-)critical temperature for chiral symmetry restoration: $T_c = 154 \pm 8$ (stat.) $\pm 1$ (sys.) MeV \cite{Bazavov:2011nk} vs. $T_c = 155 \pm 3$ (stat.) $\pm 3$ (sys.) MeV \cite{Borsanyi:2010bp}. Despite this agreement, systematic uncertainties remain in the interaction measure. Results from the HotQCD collaboration agree well with those from the Wuppertal-Budapest collaboration below $T_c$, but are systematically larger by approximately $25\%$ above $T_c$.

The production of thermal photons and dileptons is potentially more sensitive to the equation of state, because they are produced throughout the evolution and hence probe the dynamics of the system more directly \cite{Gale:2009gc,Vujanovic:2012nq,Kapusta:2006pm}.
Another relevant aspect is the inclusion of chemical freeze-out in the equation of state (and the freeze-out procedure) to reproduce the correct final particle ratios \cite{Hirano:2002ds,Kolb:2002ve,Shen:2010uy}. 

The determination of the equation of state from lattice QCD has thus far been restricted to zero baryon chemical potential, which is a good approximation
at LHC and top RHIC energies at mid-rapidity. However, away from mid-rapidity and at smaller energies the finite net-baryon density becomes an important factor.  Because lattice QCD is plagued by the “sign-problem” at non-zero baryon chemical potential, which prohibits the use of standard Monte Carlo techniques, the extraction of an equation of state at non-zero baryon chemical potential is very challenging. However, at small values of the baryon chemical potential, the construction of a more general equation of state using a Taylor expansion of the pressure \cite{Allton:2005gk} is possible.
Recently, such equation of state for finite net-baryon densities has been parametrized \cite{Huovinen:2012xm} using the results of a lattice QCD analysis of the Taylor coefficients \cite{Miao:2008sz,Cheng:2008zh}.

\section{Bulk viscosity and limits of second order viscous hydrodynamics}
QCD is a non-conformal theory and thus contains a finite bulk viscosity. The effect of bulk viscosity on elliptic flow \cite{Denicol:2009am} 
and the combined effect of shear and bulk viscosity \cite{Monnai:2009ad,Song:2009rh} have been studied. Bulk viscosity is expected to peak (possibly along with its relaxation time) around the critical temperature $T_c$, where the system may develop large correlation lengths \cite{Paech:2006st,Kharzeev:2007wb,Meyer:2009jp,Buchel:2009hv}.
Within the range of applicability of second order viscous hydrodynamics,
where viscous corrections to the thermal equilibrium distribution function remain small compared to the equilibrium part, 
corrections from bulk viscosity are found to be small compared to those from shear viscosity if the bulk relaxation time peaks around $T_c$ \cite{Song:2009rh}. This means that if bulk and shear viscous corrections do not become so large as to render the hydrodynamic expansion invalid for the relevant part of a heavy-ion collision, the extraction of the shear viscosity should be possible to reasonable accuracy when neglecting the effect of bulk viscosity. 

Recent studies of the bulk viscosity have observed that while the bulk viscosity itself scales as the second power of conformality breaking 
$\zeta \sim \eta(c_s^2-1/3)^2$, the correction to the distribution function $\delta f$ scales as the first power. \cite{Dusling:2011fd}
Corrections to the spectra are therefore dominated by viscous corrections to the distribution function, and reliable bounds on the bulk viscosity require accurate calculations of $\delta f$ in the hadronic phase. We will return to this point when discussing viscous corrections to the distribution function in Section \ref{sec:viscousCorrections}. 
However, also in this study the effect of bulk viscosity on the $p_T$-integrated pion $v_2$ at RHIC energy is found to be small.

Apart from viscous corrections to the thermal equilibrium distribution functions becoming large, bulk and shear viscous corrections can lead to a negative longitudinal pressure, argued to potentially cause the breakup of the system into droplets \cite{Torrieri:2008ip,Rajagopal:2009yw,Bhatt:2011kr}.
However, second order viscous hydrodynamics can only hint at when such cavitation could happen but is not suited to describe the process, as it happens outside the range of its applicability.

\section{Progress beyond second order viscous hydrodynamics}\label{sec:beyond2nd}
The problem with the viscous correction to the stress energy tensor becoming larger than the equilibrium part, leading to negative pressure,
arises especially in the early stage of the evolution, when the local momentum distribution is not yet equilibrated but highly anisotropic due to rapid longitudinal expansion. Recent progress has been made in describing this early time evolution and late time hydrodynamics within the same framework by performing the hydrodynamic expansion around an anisotropic distribution \cite{Martinez:2010sc,Martinez:2010sd,Florkowski:2010cf,Ryblewski:2011aq}. 
This results in new equations of motion, including one for the degree of anisotropy of the distribution function. 
The procedure can reproduce both the limits of free streaming and ideal hydrodynamics and results in second order viscous hydrodynamics 
when expanding around a small anisotropy parameter, which so far has been shown in the one dimensional \cite{Martinez:2010sc} and boost-invariant 2+1 dimensional cases \cite{Martinez:2012tu}. Another interesting aspect of this anisotropic hydrodynamics is the fact that entropy production is modified as compared to second order viscous hydrodynamics. One would expect an increase in entropy production up to some value of viscosity over entropy density followed by a decrease as one approaches the limit of free streaming. This behavior is reproduced in anisotropic hydrodynamics \cite{Martinez:2010sc} but missed in 
second order viscous hydrodynamics as the free streaming limit is far outside its range of applicability.

Another logical step is to expand to third order in gradients \cite{El:2009vj}. In the third order theory, numerical differences to second order Israel Stewart theory
are completely negligible for $\eta/s=0.05$ but become significant for $\eta/s\gtrsim 0.2$. It has also been pointed out that the hydrodynamic equations depend on the details of their derivation. While Israel and Stewart used the second moment of the Boltzmann equation to derive hydrodynamic equations for the dissipative currents, in recent work \cite{Denicol:2010xn} the definition of the latter was used directly. This leads to equations of motion of the same form but with different coefficients. Microscopic transport calculations \cite{Xu:2004mz,Xu:2007ns} show very good agreement with solutions to these equations of motion up to $\eta/s\sim 3$, while the Israel-Stewart equations show large differences for $\eta/s\gtrsim 0.2$. 
Furthermore, solutions to the Israel-Stewart formalism for heat flow do not agree with direct numerical solutions of the Boltzmann equation \cite{Bouras:2010hm}, even when the Knudsen number is very small (close to the ideal case).
An improved derivation of relativistic dissipative fluid-dynamics from kinetic theory without using the 14-moment approximation
was recently presented \cite{Denicol:2012cn,Denicol:2012vq} and provides a good description (compared to direct numerical solutions of the Boltzmann equation) of all dissipative phenomena, including heat flow. 
This demonstrates that the details of the derivation of the system of relativistic dissipative fluid-dynamic equations are relevant in particular when matching to kinetic theory at late times, as done in simulations using so called hadronic afterburners (see Section \ref{sec:after}).

\section{Temperature dependent $\eta/s$}
By comparison of experimental data with viscous hydrodynamic calculations using constant $\eta/s$ we can extract at best an effective $\langle \eta/s \rangle$. In reality $\eta/s$ should depend on the local temperature of the medium, dropping from large values at high temperatures to a minimum around $T_c$, and rising again with decreasing temperature in the hadronic phase \cite{Nakamura:2004sy,Demir:2008tr,NoronhaHostler:2008ju,Denicol:2010tr,Niemi:2011ix}.

In fact, calculations with constant $\eta/s$ find good agreement with experimental data using a smaller effective value of $\eta/s$ at RHIC than at LHC \cite{Song:2011qa,Gale:2012rq}, in line with the expected increase of $(\eta/s)(T)$ at high temperatures.

It has been shown that when including a modeled temperature dependence in calculations at RHIC energies, the details of $\eta/s(T)$ in the quark-gluon plasma phase have little influence on the final elliptic flow result, while hadronic $\eta/s(T)$ modifies $v_2$ strongly \cite{Niemi:2011ix}. At the highest LHC energies the conclusion is the opposite: weak dependence on the hadronic $\eta/s$ but strong dependence on $\eta/s(T)$ in the QGP phase. Interestingly, at RHIC energies, a significant dependence has been found on the minimum value of $\eta/s(T)$ around $T_c$ \cite{Niemi:2012ry}.

In addition, a strong dependence of the entropy production on the initial value of $\pi^{\mu\nu}$ was found in \cite{Shen:2011eg} when starting with a large $\eta/s$. After correcting for this to reproduce final particle spectra, differences in $v_2$ were however minor. Yet, particularly when including a temperature dependent $\eta/s$, it is essential to gain a better understanding of the pre-equilibrium stage in heavy-ion collisions to determine the initial conditions for viscous hydrodynamics.

Recent simulations have also incorporated a temperature dependent $(\eta/s)(T)$  \cite{Bozek:2011ua,Gale:2012rq} and bulk viscosity over entropy density ratio $(\zeta/s)(T)$ \cite{Bozek:2011ua} in 3+1 dimensional simulations.

\section{Conversion into particle spectra}
To compare to experimental measurements, the evolved stress-energy tensor has to be converted into particle spectra.
It is possible to perform the kinetic freeze out in the hydrodynamic simulation, given a freeze-out condition.
The typically used condition is the system reaching a certain freeze-out temperature or energy density. However, it has been pointed out
\cite{Hung:1997du,Holopainen:2012id} that a realistic description should involve a comparison of the system's expansion rate with the particle
scattering rate: when the scattering rate is not significantly larger than the expansion rate the system will freeze out.

Particle spectra can be computed using the Cooper-Frye formula \cite{Cooper:1974mv}
\begin{equation}\label{eq:CF}
  E\frac{dN}{d^3p} = \int_\Sigma d\Sigma_\mu p^\mu f(T,p_\mu u^\mu,\pi^{\mu\nu}) \,,
\end{equation}
where $ f(T,p_\mu u^\mu,\pi^{\mu\nu}) $ is the particle distribution function, which is thermal in the ideal case and includes corrections
proportional to $\pi^{\mu\nu}$ in the viscous case. $\Sigma$ is the freeze-out surface, generally a three-dimensional surface in 3+1 dimensional space-time,
which characterizes the distribution of time and space where the system freezes out.

Another, perhaps more realistic way to deal with the late stage of a heavy-ion collisions when the system becomes dilute is to switch from a 
hydrodynamic description to a hadronic cascade simulation \cite{Bass:2000ib,Teaney:2000cw,Hirano:2005xf,Nonaka:2006yn,Nonaka:2010zz,Hirano:2010jg,Hirano:2010je}. Conversion into individual particles can be done by sampling the spectra determined from (\ref{eq:CF}). 
These particles are then evolved in the hadronic cascade and freeze-out happens automatically as the system continues to expand and interactions
become very rare. 

We discuss several important aspects of the conversion of hydrodynamic quantities into measurable particles in the remainder of this section.

\subsection{Viscous corrections to particle distribution functions}\label{sec:viscousCorrections}
As alluded to above, when translating the dissipative $T^{\mu\nu}$ to particles in the Cooper-Frye formalism \cite{Cooper:1974mv}, corrections to the distribution function $\delta f$ have to be taken into account:
\begin{equation}
 T^{\mu\nu}_{\rm hydro} = \sum_{n=1}^N d_n \int \frac{d^3p}{(2\pi)^3}\frac{p^\mu p^\nu}{E_n} (f_{0n}+\delta f_n) 
\end{equation}
for an $N$ component system with $d_n$ the degeneracy of species $n$.
 The ansatz
\begin{equation}
  \delta f_n = \frac{C_n}{2T^3}f_0 (1\pm f_0) \hat{p}^\alpha \hat{p}^\beta \chi(p) \frac{\pi_{\alpha\beta}}{\eta}\,,
\end{equation}
where $\hat{p}^\alpha$ is a unit vector in the $\alpha$ direction, leaves some freedom and the usual procedure is to assume that all coefficients $C_n$ are equal, even though they should depend on the individual particle species' interaction rate \cite{Molnar:2011kx}, and use $\chi(p)=p^2$, which is derived within a relaxation time plus Boltzmann approximation:
\begin{equation}
  \delta f_n = f_{0n} (1\pm f_{0n}) p^\alpha p^\beta \pi_{\alpha\beta} \frac{1}{2 (\epsilon+\mathcal{P}) T^2} ~ \forall ~  n\,.
\end{equation}
It has been shown \cite{Dusling:2009df} that $\chi(p)\propto p^\alpha$, where $\alpha$ can take on values from 1 to 2, which is the case for example for a system with radiative and elastic energy loss that has $\chi(p)\propto p^{1.38}$. The exact form of the correction has a sizable effect on the $p_T$ differential elliptic flow for $p_T\gtrsim 1\,{\rm GeV}$. It should be noted, however, that the analysis of experimental data \cite{Lacey:2010fe} indicates $\chi(p)\propto p^2$.
A more general problem is that corrections $\delta f$ can become large compared to $f_0$. For hadrons this problem can be reduced when using a hadronic afterburner and switching at intermediate temperatures of $\sim 160\,{\rm MeV}$, but for photons that are produced throughout the whole evolution this becomes a serious concern \cite{Dion:2011pp}.

As mentioned earlier, corrections from bulk viscosity are most prominent in modifications of $\delta f$, which is proportional
to the conformal breaking. Therefore accurate calculations of $\delta f$ in the hadron resonance gas are required. \cite{Dusling:2011fd}
While effects of bulk viscosity on integrated $v_2$ seem small, bulk viscous effects need more detailed studies to determine $\zeta/s$
and $\eta/s$ as functions of temperature in the future.

\subsection{Hadronic afterburner}\label{sec:after}

Hydrodynamic description of a system applies when the mean free paths of
constituent particles are small compared to any macroscopic length scale.
It is then well justified to use
hydrodynamics for the initial phase of the QGP evolution since the density
is so high.
As the system expands and cools, the mean free paths decrease and the ratio
$\eta/s$ increases.
Eventually, when the density becomes low enough,
the fact that each species has a different mean
free path begins to matter.
This is because differing mean free path not only 
implies different speed of response to the change in the collective flow,
but it also implies that the collective description is becoming
inadequate.
On the other hand, the kinetic theory model of heavy ion collisions works
best when the system is dilute and the mean free paths long. Furthermore,
it is only natural that the mean free path of each species is different in
the kinetic theory models.

If one can assume that there is a range of temperature and density that the
validity of both the hydrodynamic approach and the kinetic theory approach
overlap, then it becomes natural that the
hydrodynamic description 
should give way to a kinetic theory description
when temperature and density fall within this range.
Although it is not rigorously proven that such a transition region exists,
evidence shows that including such transition in the simulation
does improve the agreement between theory and experiments\cite{Bass:2000ib,Teaney:2000cw,Nonaka:2006yn,
Petersen:2008dd,Song:2010aq,Song:2010mg,Song:2011hk,Song:2011qa,Shen:2011zc},
especially in the baryonic sector.

Coupling between an event-by-event 3+1D viscous hydrodynamics 
and a kinetic theory model has been recently accomplished\cite{Ryu:2012at}.
The publicly available UrQMD code (v.~3.3p1) 
\cite{Petersen:2008dd} was joined with the event-by-event 3+1D viscous hydrodynamic simulation \textsc{music},
taking special care in matching the $\eta-\tau$ coordinate system used in \textsc{music} to the $t-z$
system used in UrQMD. 
To couple \textsc{music} to UrQMD, first the isothermal 3-D hyper-surface at $T= 170\,\hbox{MeV}$ is determined in the hydrodynamic simulation.
Next the transition hyper-surface is sampled using the Cooper-Frye formula Eq.\,(\ref{eq:CF})
to populate hadrons in $t-z$ space. These hadrons are then propagated backward in time
without collisions to the common UrQMD initial time $t_0$.

At this point, the regular UrQMD evolution takes over, but with the important difference that the
particles are allowed to interact only {\em after} they emerge from the isothermal hyper-surface.
This is done because different cells reach the transition temperature
at different times whereas UrQMD needs to evolve particles in Minkowski time.
In this way, particles are simulated as emerging from the plasma at different times.

\begin{figure}[t]
\centerline{
\includegraphics[width=0.45\textwidth,height=4cm]{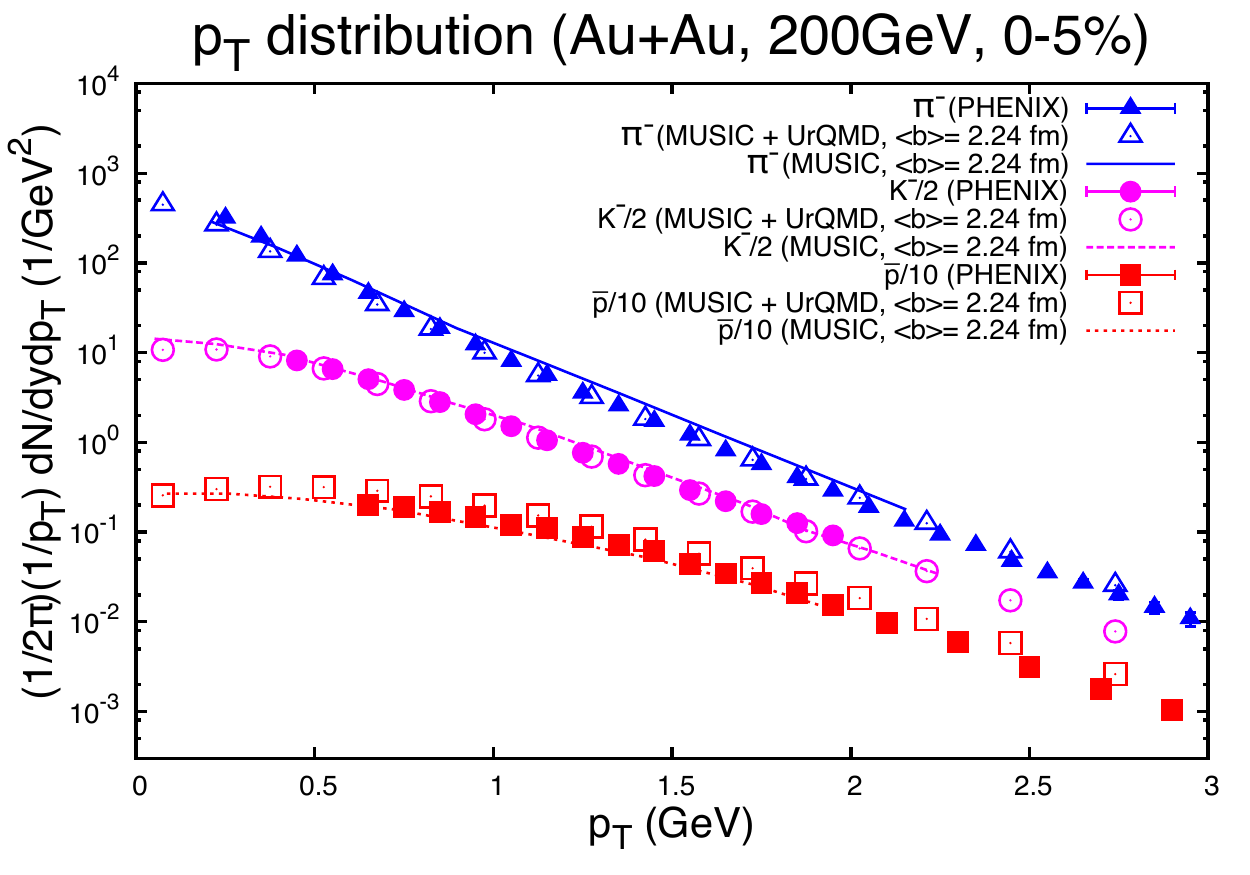}
\includegraphics[width=0.45\textwidth,height=4cm]{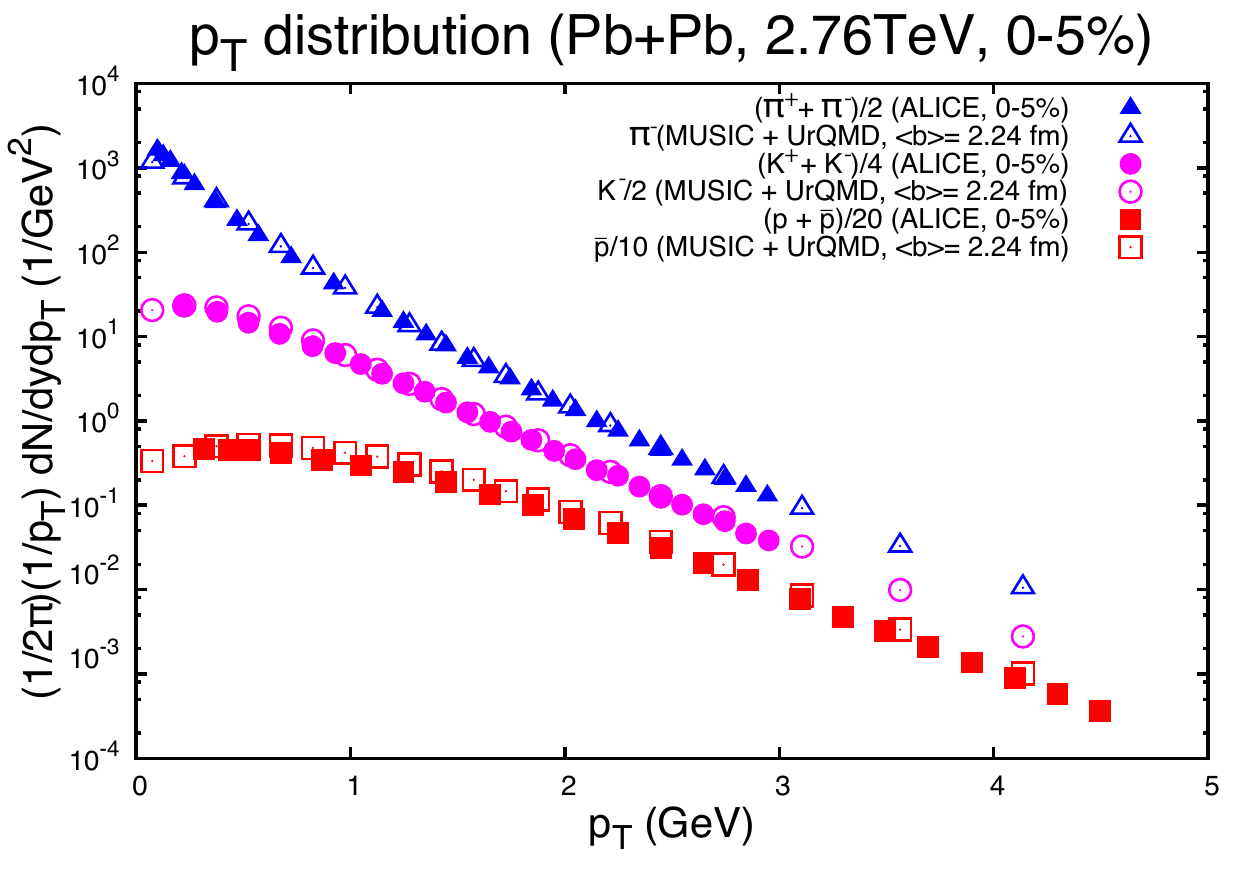}
}
\caption{
Transverse momentum distribution of identified particles at RHIC and LHC for 0-5\,\%
most central collisions compared to \textsc{music}$+$UrQMD calculations. RHIC result is based on 10,000
events (100 UrQMD events on each of 100 hydro events).
LHC result is based on 1,000 events (10 UrQMD events on each of 100 hydro events).
For RHIC, the results of pure \textsc{music} events are also shown. Experimental data from PHENIX \cite{Adler:2003cb} and ALICE (preliminary) \cite{Milano:2012qm}.
}
\label{fig:dNdpT}
\end{figure}

\begin{figure}[t]
\centerline{
\includegraphics[width=0.45\textwidth,height=4cm]{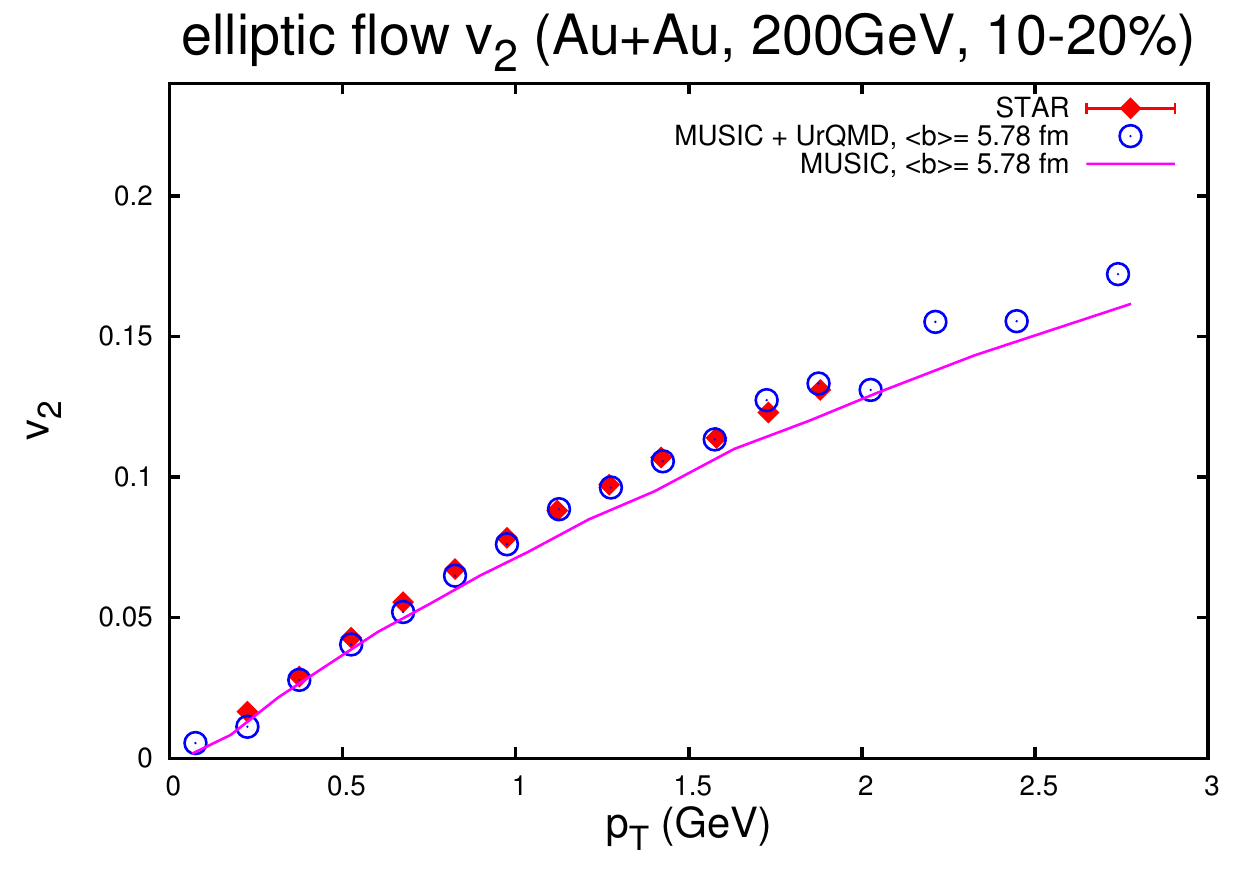}
\includegraphics[width=0.45\textwidth,height=4cm]{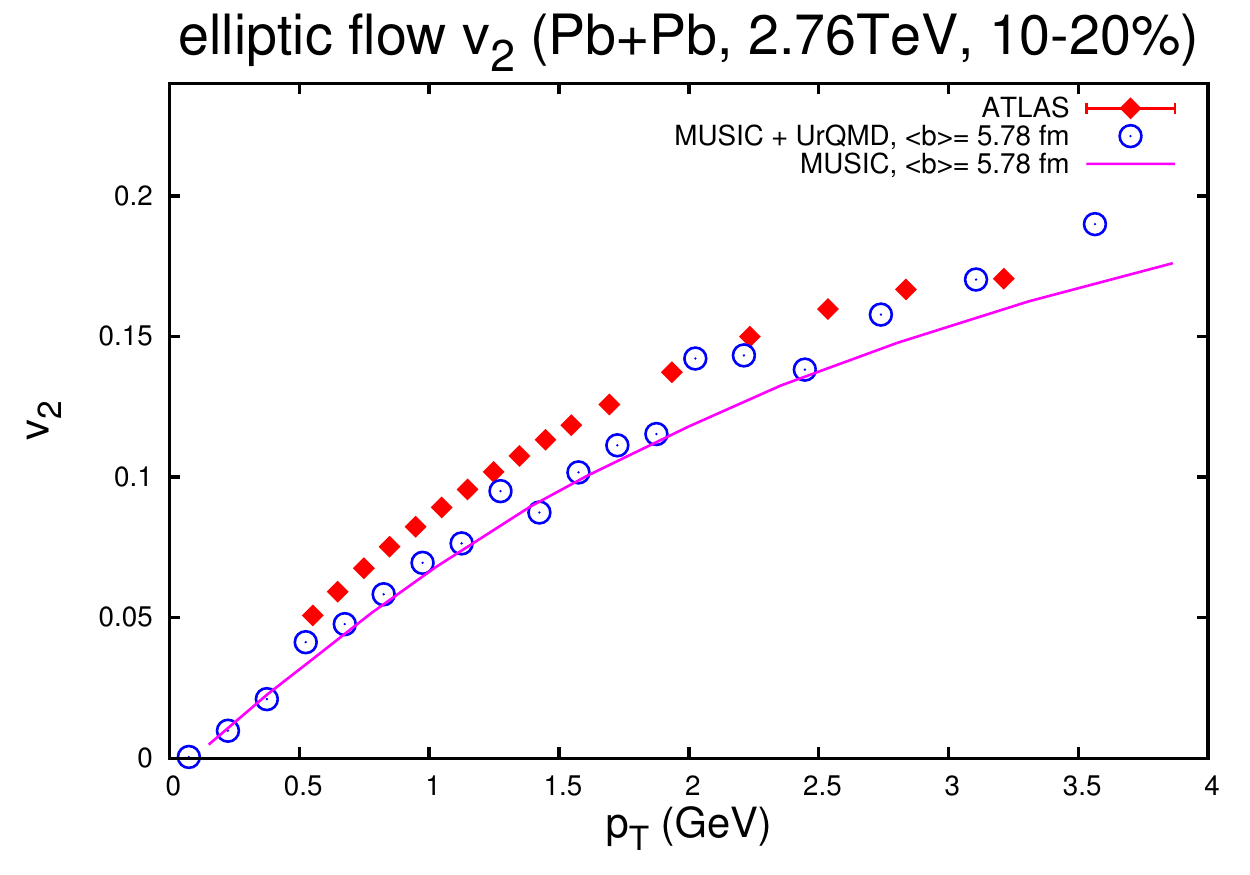}
}
\caption{
The $p_T$ dependent
$v_2$ from RHIC and the LHC in the $10 - 20\,\%$
centrality class
compared to \textsc{music} and \textsc{music}$+$UrQMD calculations. 
RHIC result is based on 10,000 events
and LHC result is based on 1,000 events. Experimental data from STAR \cite{Adams:2004bi} and ATLAS \cite{ATLAS:2011ah}.
}
\label{fig:v2}
\end{figure}

\begin{figure}[t]
\centerline{
\includegraphics[width=0.45\textwidth,height=4cm]{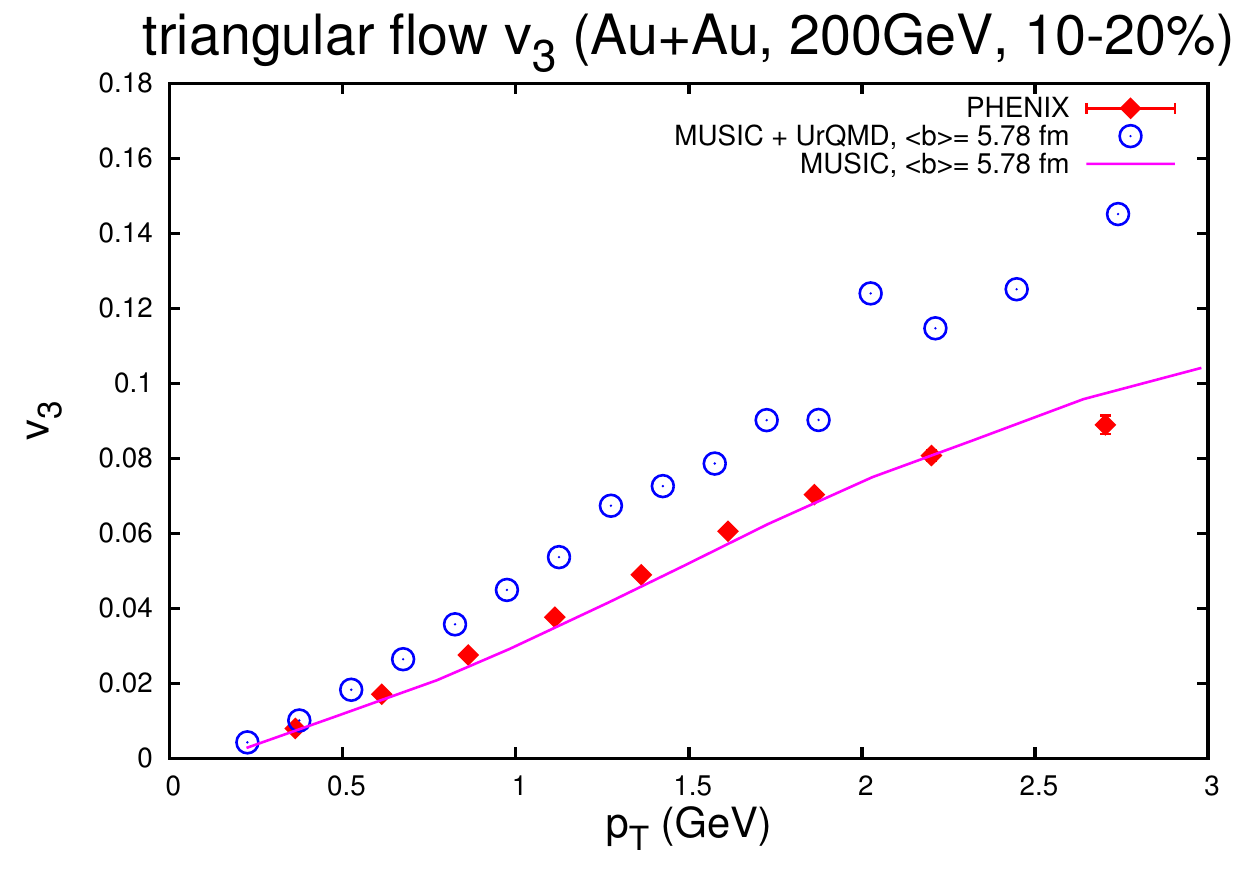}
\includegraphics[width=0.45\textwidth,height=4cm]{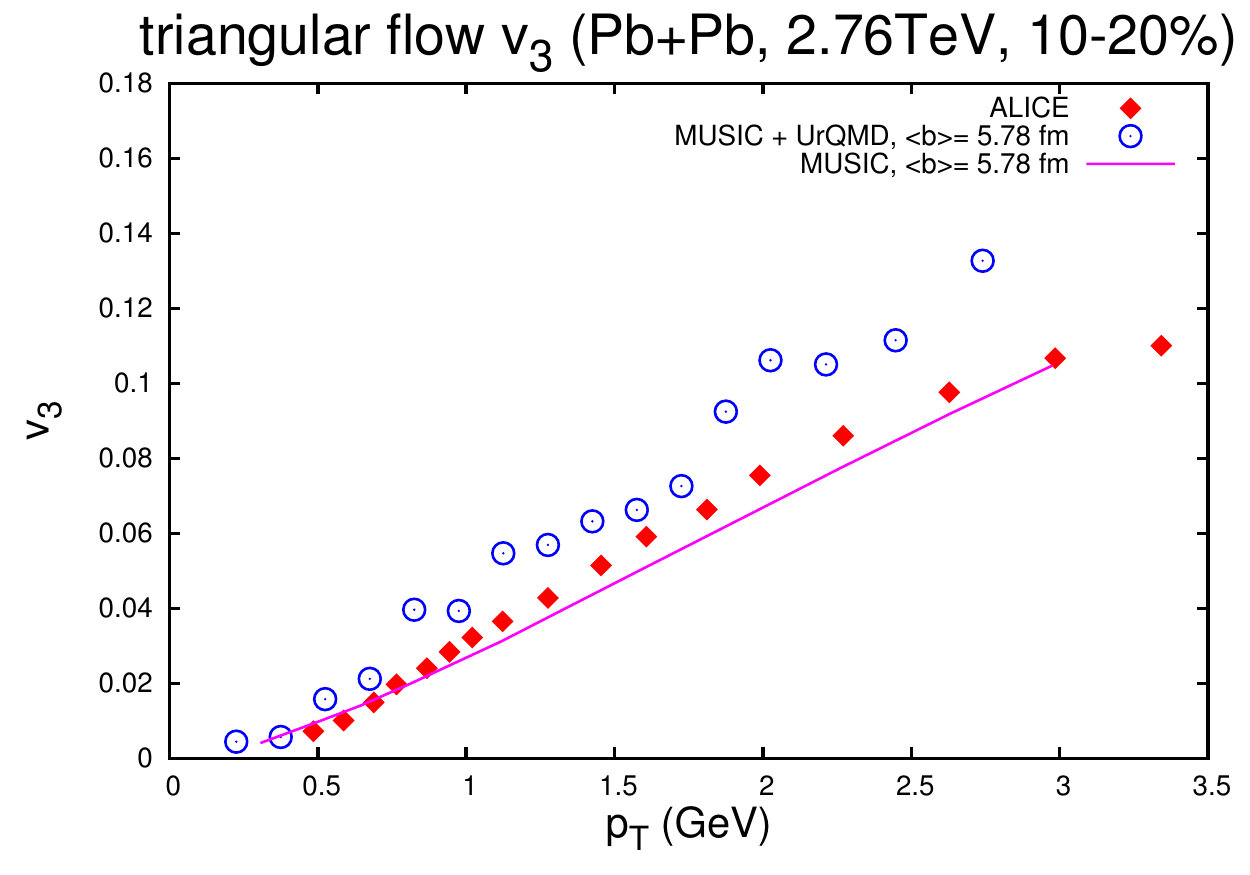}
}
\caption{
The $p_T$ dependent
triangular flow from RHIC and the LHC in the $10 - 20\,\%$
centrality class
compared to the \textsc{music} and \textsc{music}$+$UrQMD calculations. 
RHIC result is based on 10,000
events
and LHC result is based on 1,000 events. 
Experimental data from PHENIX \cite{Richardson:2012kq} 
and ALICE \cite{Abelev:2012di}.
}
\label{fig:v3}
\end{figure}

Three main results from this study
are shown in Figs. \ref{fig:dNdpT},\ref{fig:v2}, and \ref{fig:v3}.
Hybrid calculations with microscopic cascade simulations typically use oversampling
to save computing time.
In the case of \textsc{music} with UrQMD in each centrality bin, 100 \textsc{music} events are sampled 
100 times each for RHIC and 1000 \textsc{music} events are sampled 10 times each for LHC.
Light hadron spectra for $\pi^-$, $K^-$ and $\bar{p}$ for RHIC are shown in the left panel of Fig. \ref{fig:dNdpT}.
Experimental data from PHENIX \cite{Adler:2003cb} is reasonably well reproduced for the shown $0 - 5\,\%$ centrality bin. 
For all three species, the description of the data is either 
improved over or of about the same quality as the pure hydrodynamics results.
In the second panel, LHC predictions are compared with the ALICE data shown at QM2012 \cite{Milano:2012qm}.

For $v_2(p_T)$ shown in Fig. \ref{fig:v2}, the $\bar{p}$ result is much improved with the afterburner at RHIC. 
This may be due to the better description of the finite baryon mean free path in UrQMD compared to hydrodynamics.
For the LHC, one may argue that the afterburner improves the description a little, 
but the statistical error at this point is too big to make a definite statement.
Calculations with afterburner lead to larger $v_3(p_T)$ in general (see Fig. \ref{fig:v3}). However, this needs more careful study to quantify. 

%


\section{Event-by-event hydrodynamics}
\subsection{Overview}
It is now generally accepted that fluctuating initial conditions for hydrodynamic simulations of heavy-ion collisions
are essential for the exact determination of collective flow observables and to describe 
features of multi-particle correlation measurements in heavy-ion collisions
\cite{Andrade:2006yh,Adare:2008cqb,Abelev:2008nda,Alver:2008gk,Alver:2009id,Abelev:2009qa,Miller:2003kd,Broniowski:2007ft,Andrade:2008xh,Hirano:2009bd,
Takahashi:2009na,Andrade:2009em,Alver:2010gr,Werner:2010aa,Holopainen:2010gz,Alver:2010dn,Petersen:2010cw,Schenke:2010rr,Schenke:2011tv,Qiu:2011iv}.
Real event-by-event hydrodynamic simulations have been performed and show modifications to spectra and flow from ``single-shot'' 
hydrodynamics with averaged initial conditions \cite{Holopainen:2010gz,Schenke:2010rr,Schenke:2011tv,Qiu:2011iv}.
An important advantage of event-by-event hydrodynamic calculations is the possibility to study higher flow harmonics such as $v_3$, 
which is entirely due to fluctuations like all odd harmonics. 
Different $v_n$ depend differently on $\eta/s$ and the details of the initial condition, 
which is determined by the dynamics and fluctuations of partons in the incoming nuclear wave functions.
This observation can be used to determine these long sought after
details of the initial state and medium properties in heavy-ion collisions by performing a systematic analysis
of all harmonics $v_n$, up to e.g. $n=6$ as a function of $\eta/s$ and the initial state properties and compare to experimental data.
First predictions of $v_3$ from hydrodynamic simulations \cite{Schenke:2010rr} agree extremely well with experimental data from RHIC \cite{Adare:2011tg}.
Furthermore, it has been shown that at low $p_T$ (and $|\Delta \eta|>1$ (ALICE), $|\Delta \eta|>2$ (ATLAS)), the main features 
of dihadron correlations in the angular difference $\Delta \phi$ between the hadron momenta can be described by flow, i.e.,
 the sum of $v_1$ to $v_6$ only \cite{Jia:2011hfa,ALICE:2011vk}. 
The double-peak structure on the away-side is hence described mostly by (triangular) flow as some works had predicted \cite{Takahashi:2009na,Alver:2010gr}.
Detailed measurements and calculations of new quantities, such as event plane correlations or event-by-event distributions of flow coefficients
have appeared over the last year and will be reviewed in this section.

\subsection{Initial state models}
There are several ways for describing and generating fluctuating initial conditions for hydrodynamics. Generally, this means the energy density
distribution at the ``thermalization time'' and in some cases the distribution of flow velocities.
One of the most commonly used models is the Monte-Carlo (MC) Glauber model \cite{Miller:2007ri}. 
In its simplest implementation uncorrelated nucleons are sampled from measured density distributions. 
Then the two nuclei are arranged according to a random impact parameter $b$ and projected onto the transverse $x$-$y$ plane, assuming straight line trajectories for all nucleons. 
Interaction probabilities are then computed using the relative distance between two nuclei and the measured nucleon-nucleon inelastic cross section.
Every wounded nucleon (sometimes also a fraction of binary collision points) is then assigned an energy or entropy density, parametrized e.g. 
as a two dimensional Gaussian in the transverse plane.
The model can be improved by the inclusion of many-body correlations between the nucleons. However, it was shown that the inclusion of realistically 
correlated configurations does not modify the geometry, characterized by the dipole asymmetry and triangularity of the distribution, 
compared to the uncorrelated case \cite{Alvioli:2011sk}.

The MC-KLN (Monte-Carlo Kharzeev-Levin-Nardi) model, another commonly used initial state model, is based on the color-glass-condensate framework.
Within this framework, one takes into account the well established feature of QCD that at small Bjorken-$x$, a novel regime governed by large gluon 
densities and non-linear coherence phenomena takes over \cite{Iancu:2003xm}. These high gluon densities correspond to strong classical fields, permitting calculations of the wave function using classical techniques. Quantum corrections are then incorporated via non-linear renormalization group equations 
such as the JIMWLK  or in the large $N_c$ limit the BK equations \cite{Balitsky:1995ub,Kovchegov:1999yj} that describe the evolution of the wavefunction towards smaller $x$.
Certain saturation models incorporate this evolution via parametrizations of the unintegrated gluon distribution or other quantities and we 
describe some of these approaches in detail below.

The MC-KLN model in particular is built upon the KLN model \cite{Kharzeev:2000ph,Kharzeev:2001gp}, which was improved to the so called f(actorized)KLN model and extended to a full Monte-Carlo version \cite{Drescher:2006ca,Drescher:2007ax}. 
The main ingredient for this description is the $k_T$-factorized expression for the gluon multiplicity, essentially a convolution of the unintegrated
gluon distributions $\phi_{A/B}$ of the two colliding nuclei:
\begin{align}\label{eq:KLN}
  \frac{dN_g}{d^2r_T dy} = \frac{4 N_c}{N_c^2-1}\int^{p_T^{\rm max}}&\frac{d^2p_T}{p_T^2}\int^{p_T}\frac{d^2k_T}{4}\alpha_s \nonumber\\
  & \times \phi_A(x_1,(\mathbf{p}_T+\mathbf{k}_T)^2/4) \phi_B(x_2,(\mathbf{p}_T-\mathbf{k}_T)^2/4)
\end{align}
Here, $p_T$ and $y$ denote the transverse momentum and the rapidity of the produced gluons, respectively. $\mathbf{r}_T$ is the position in the transverse plane. The light-cone momentum fractions of the colliding gluon ladders are then given by $x_{1,2}=p_T \exp(\pm y)/\sqrt{s}$, where $s$
denotes the center of mass energy. 

A few remarks on $k_T$-factorization are in order. A general shortcoming of $k_T$-factorization is that multiple scatterings for $p_T < Q_s$ are omitted even though they should be present because occupation numbers are high. So the approach is strictly not valid in the fully non-linear case of A+A collisions \cite{Blaizot:2008yb}. In fact, Eq.(\ref{eq:KLN}) cannot be derived for collisions of two dense systems but is rather an extrapolation from the result in the ``dilute'' p+A limit. The cutoff of the $k_T$ integral at $p_T$, used in the MC-KLN model, is introduced by hand to make the spectrum infrared finite. 
Different implementations of $k_T$-factorization and comparisons to more appropriate classical Yang-Mills calculations have been discussed in the literature \cite{Blaizot:2010kh}.

The KLN approach uses a particular parametrization of the unintegrated gluon distribution,
depending on the saturation scale $Q_s^{A/B}$ \cite{Kharzeev:2000ph,Kharzeev:2001gp}.
Apart from the mentioned problems with $k_T$-factorization for A+A collisions, there are several problems with the original KLN approach, in particular that the saturation scale in one nucleus depends on properties of both nuclei \cite{Lappi:2006xc}, and that the limit at the edge of a nucleus does not approach the unintegrated gluon distribution of a single nucleon.
These problems are cured in the fKLN approach \cite{Drescher:2006ca} where the result for the gluon multiplicity is truly factorized (unintegrated gluon distributions of nucleus A (B) only depend on properties of nucleus A (B))
\begin{align}\label{eq:fKLN}
  \frac{dN_g}{d^2r_T dy} \sim \int\frac{d^2p_T}{p_T^2}\int d^2k_T\, p_A\, p_B\,  
  \phi_A(T_A/p_A)\, \phi_B(T_B/p_B)\,,
\end{align}
where $p_{A/B}$ is the probability for finding at least one nucleon at a given transverse coordinate in nucleus $A/B$, with $T_{A/B}$ being
the thickness function.
The implementation of Monte-Carlo sampling also takes care of correct behavior at the edges.
The initial energy density is obtained by computing the transverse energy distribution of the produced gluons, i.e., by including another factor 
of $p_T$ in the integrals in Eqs. (\ref{eq:KLN},\ref{eq:fKLN}).
The model has been further extended to include running-coupling BK (Balitsky-Kovchegov) evolution of the gluon distributions 
in the rcBK model \cite{Albacete:2010ad}.
MC-KLN-models, as the MC-Glauber model, lack negative binomial fluctuations of multiplicities for a given number of participants.
In a recent extension, these fluctuations have been added to the MC-KLN model by hand \cite{Dumitru:2012yr}.  In this implementation 
the correct correlations in the transverse plane are obtained by choosing the grid cell size to be $1/Q_s$.

A more recent improved color-glass-condensate based approach is the IP-Glasma model \cite{Schenke:2012wb,Schenke:2012hg}, which combines the IP-Sat (Impact Parameter dependent Saturation Model) model~\cite{Bartels:2002cj,Kowalski:2003hm} of high energy nucleon (and nuclear) wavefunctions with the classical Yang-Mills (CYM) dynamics of the glasma fields produced in a heavy-ion collision~\cite{Kovner:1995ja,Kovchegov:1997ke,Krasnitz:1998ns,Krasnitz:1999wc,Krasnitz:2000gz,Lappi:2003bi}. 
After fixing the free parameters of the IP-Sat model by fits to small $x$ HERA deeply inelastic scattering (DIS) data off protons and fixed target nuclear DIS data~\cite{Kowalski:2006hc,Kowalski:2007rw}, the IP-Sat model provides an excellent description of these data.
The IP-Glasma model includes fluctuations of nucleon positions as well as sub-nucleonic fluctuations of color charges, a feature missing in most other initial state models. Another advantage is that the model does not rely on $k_T$-factorization, which is strictly only valid when at least one of the 
sources is dilute (as in $p+p$ and $p+A$ collisions) as discussed above. Furthermore, the IP-Glasma includes the non-linear pre-equilibrium evolution of the initial gluon fields.
This leads to the build-up of initial flow and an independence of the exact time when one switches to hydrodynamics \cite{Gale:2012rq}.
The early stage dynamics is however not fully included. Instabilities triggered by quantum fluctuations, and subsequent strong scattering of over-occupied fields, may lead to rapid isotropization and quenching of  $\Pi^{\mu\nu}$ to reasonable values justifying the use of viscous hydrodynamics already at early times.
These unstable dynamics require a full 3+1 dimensional simulation including a realistic description of quantum fluctuations, which has not yet been fully achieved. However, significant progress is being made \cite{Dusling:2011rz,Dusling:2010rm,Epelbaum:2011pc,Dusling:2012ig2,Berges:2012mc} and the IP-Glasma model can be extended to include these important effects.
Matching of the full stress-energy tensor, including viscous corrections and flow, to the hydrodynamic simulation will then be possible.
Another possibility is to couple this initial condition to anisotropic hydrodynamic simulations \cite{Martinez:2010sc,Martinez:2010sd,Florkowski:2010cf,Ryblewski:2011aq} described in Section \ref{sec:beyond2nd}.
The additional color charge fluctuations in the IP-Glasma model naturally lead to negative binomial fluctuations in the event-by-event multiplicity and the correlation length of the fluctuations in the transverse plane is of the order of the inverse saturation scale $1/Q_s$ as desired.

\begin{figure}[htb]
  \vspace{-0.5cm}
   \begin{center}
    \begin{minipage}{0.495\textwidth}
     \includegraphics[width=6.5cm]{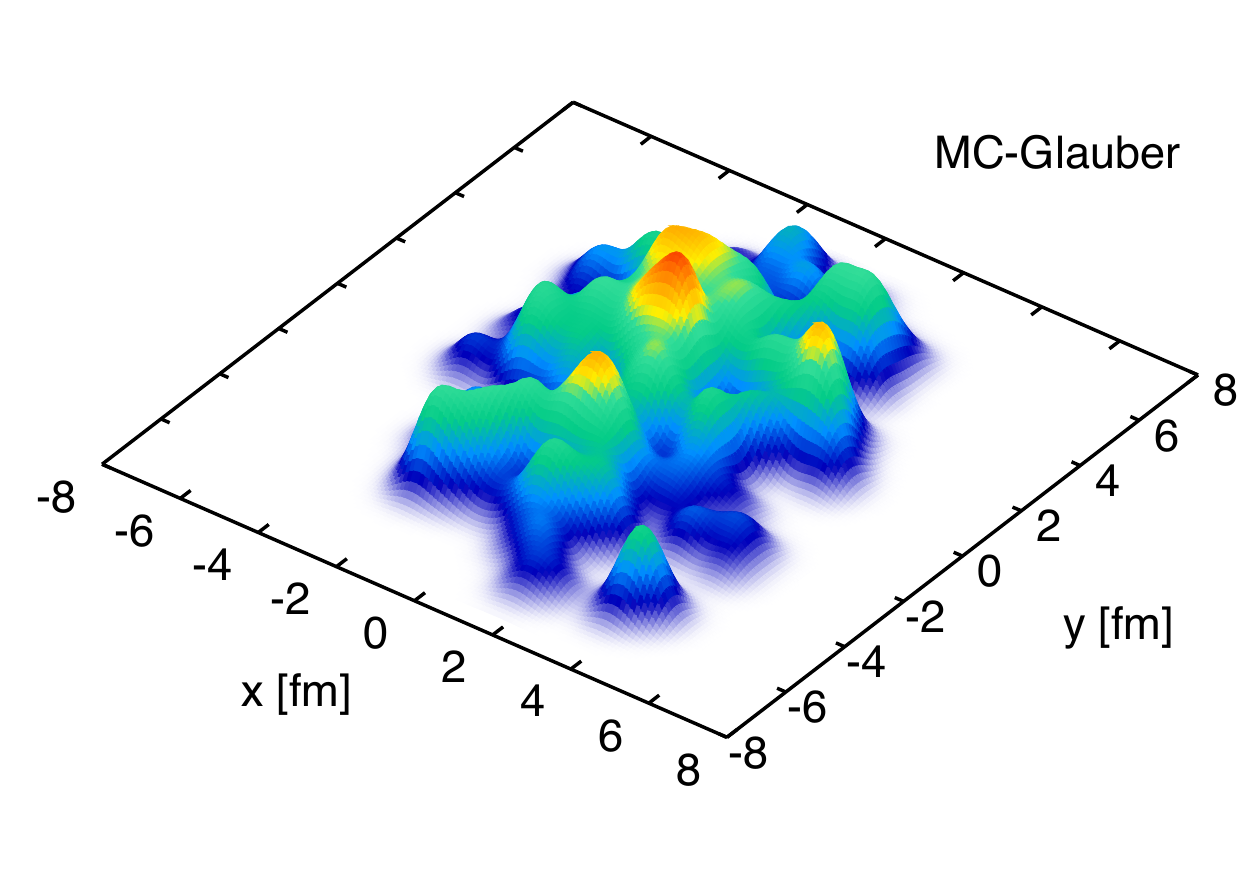}
    \end{minipage}
    \hfill
    \begin{minipage}{0.495\textwidth}
    \includegraphics[width=6.5cm]{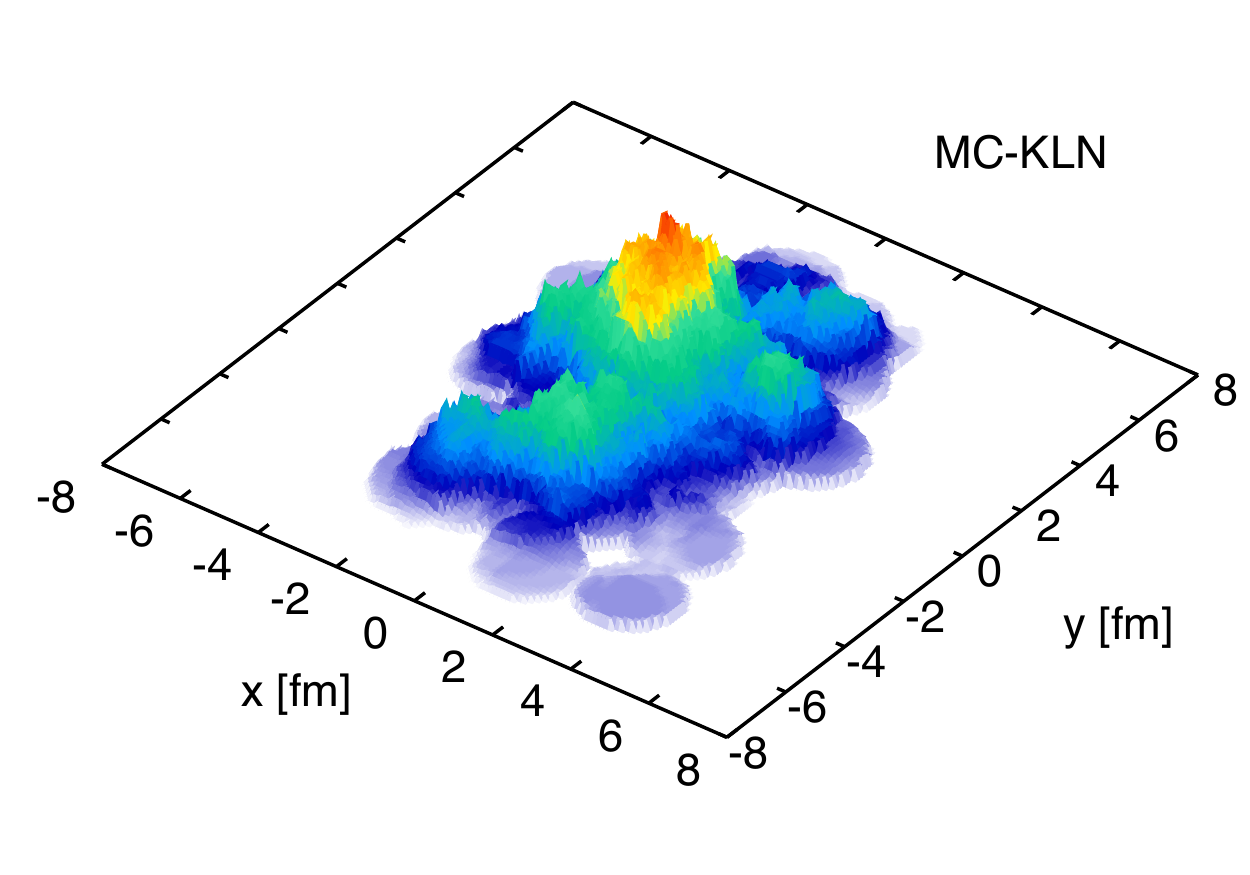}
    \end{minipage}
    \\\vspace{-0.2cm}
    \begin{minipage}{0.495\textwidth}
    \includegraphics[width=6.5cm]{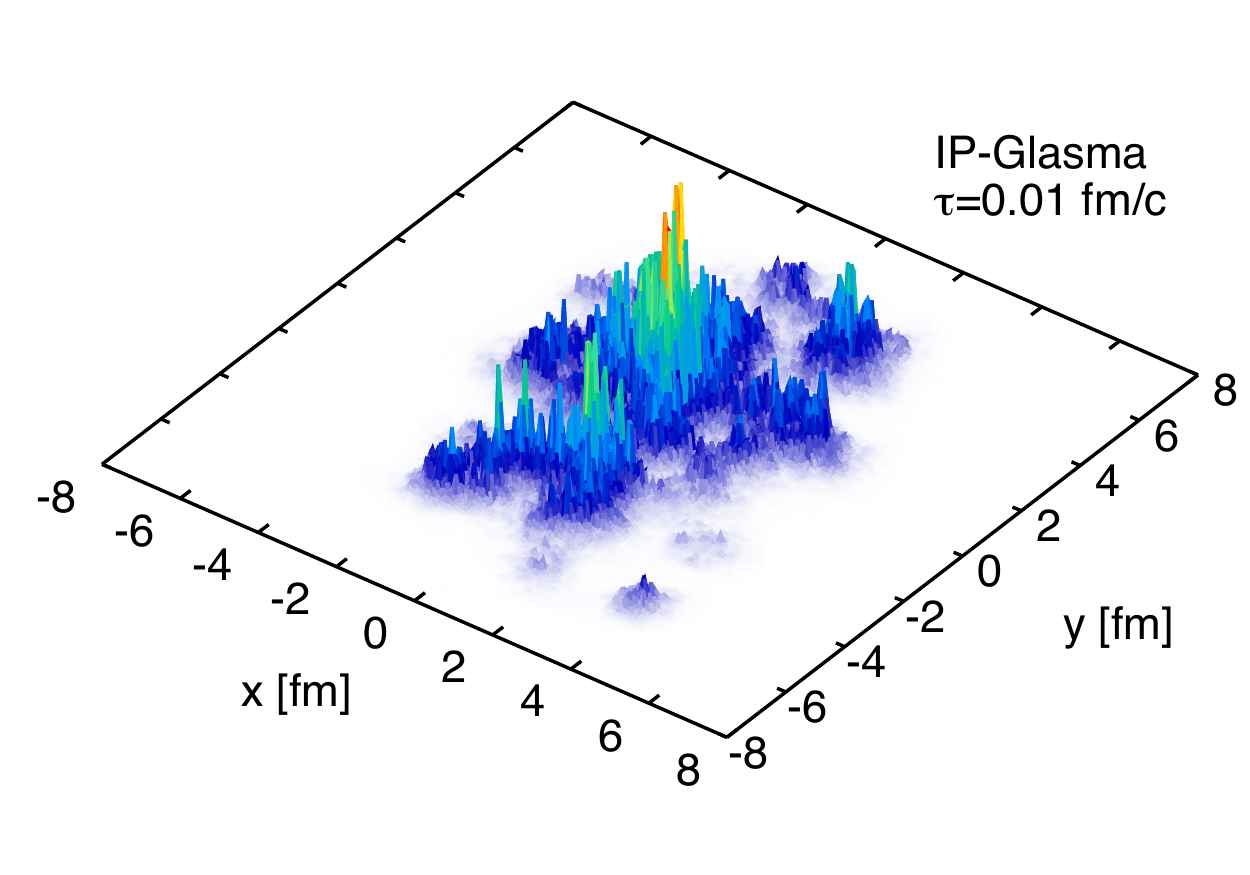}
    \end{minipage}
    \hfill
    \begin{minipage}{0.495\textwidth}
    \includegraphics[width=6.5cm]{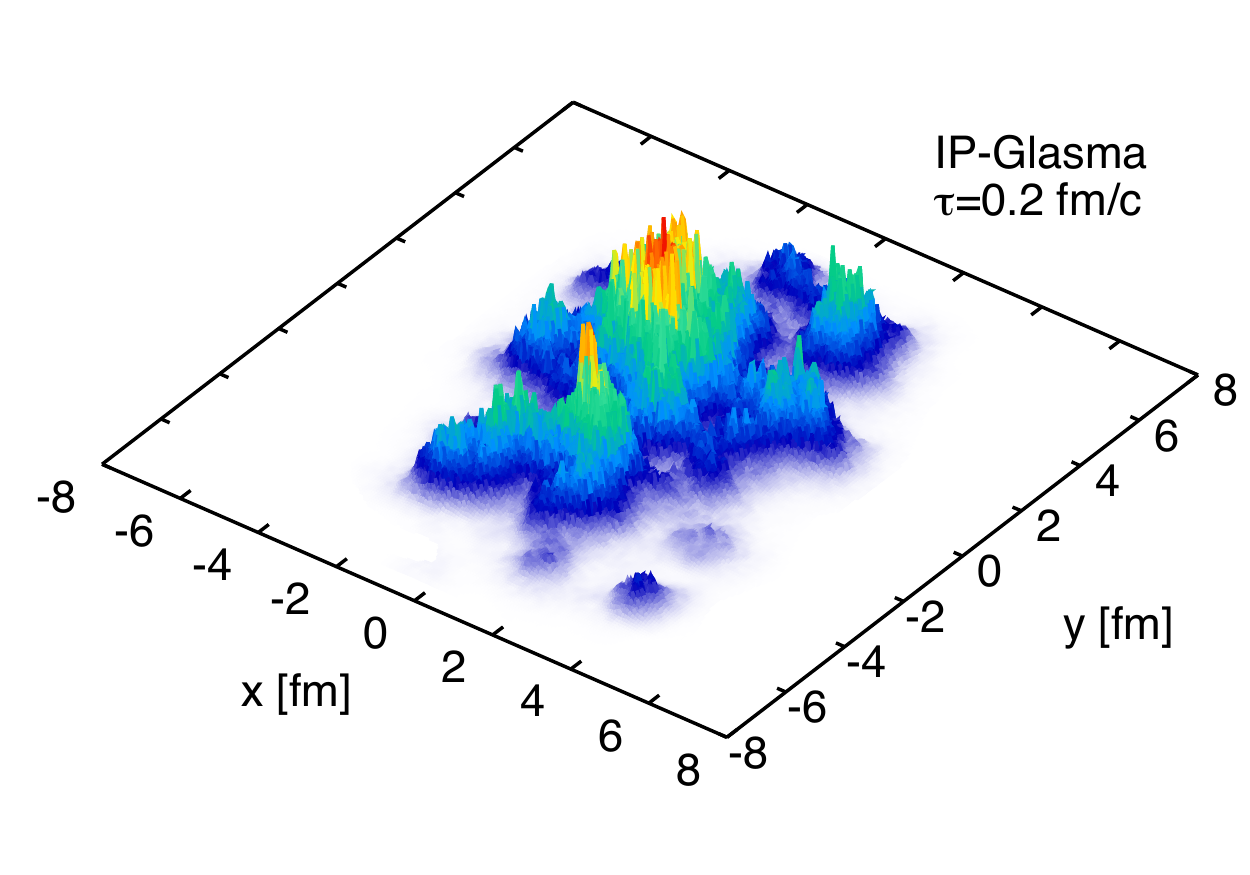}
    \end{minipage}
     \vspace{-0.5cm}
     \caption{Comparison of the initial energy density (arbitrary units) produced by the MC-Glauber, MC-KLN and IP-Glasma models. All events have the same configuration
     of nucleons and impact parameter $b=4\,{\rm fm}$ to emphasize how different model descriptions affect the structure of the energy density. The finest structure is obtained in the IP-Glasma model, which includes subnucleonic color charge fluctuations. Yang-Mills evolution to $\tau=0.2\,{\rm fm}/c$ smoothens this structure before it enters a hydrodynamic simulation. }
     \label{fig:edens}
   \end{center}
   \vspace{-0.5cm}
\end{figure}

In Fig. \ref{fig:edens} we show a comparison of initial energy densities from an MC-Glauber, the MC-KLN and the IP-Glasma model using the same distribution of nucleons in the incoming nuclei. In the MC-Glauber model every wounded nucleon was assigned a two dimensional Gaussian energy density with a width of $\sigma_0=0.4\,fm$. The MC-KLN result was obtained using the publicly available code \textsc{mckln-3.52} \cite{mckln}. IP-Glasma results are shown for two different times, $\tau=0.01\,{\rm fm}/c$ and $\tau=0.2\,{\rm fm}/c$ after Yang-Mills evolution. The evolution smoothens the initially very distinct structures noticeably. Because of the additional subnucleonic fluctuations, the IP-Glasma model produces the finest granularity, typically leading to larger fluctuation driven odd eccentricities \cite{Schenke:2012wb,Schenke:2012hg}. 

Apart from MC-Glauber and CGC based frameworks, there are several parton- and hadron-cascade models that are being used to determine fluctuating initial conditions.
These are for example UrQMD \cite{Petersen:2010cw}, EPOS \cite{Werner:2012xh}, and AMPT \cite{Pang:2012he,Pang:2012uw},
all using Monte-Carlo techniques to compute initial particle production and then converting the soft part of the spectrum into 
the bulk energy density distribution used in hydrodynamic simulations. They also provide initial flow and in principle 
the full stress-energy tensor including viscous corrections.

\subsection{Hydrodynamic fluctuations}
In addition to fluctuations in the initial state, there should be fluctuations occurring during the hydrodynamic evolution, simply due to the
finite number of particles present in a real heavy-ion collision.
These hydrodynamic fluctuations \cite{Kapusta:2011gt,Springer:2012iz} are directly related to the hydrodynamic properties of the matter, in particular, 
owing to the fluctuation-dissipation theorem, their amplitude is governed by the viscosities. This offers the possibility to constrain the viscosity of the strongly coupled quark-gluon plasma independently from the traditional analysis of anisotropic flow. Full simulations including this type of fluctuations are still to be developed.

\subsection{Numerical schemes}
When dealing with fluctuations and trying to learn about the physical viscosity of the system, one needs numerical schemes to solve the hydrodynamic
equations that can deal well with large gradients and have a small artificial viscosity. 
It has been shown that in the case of heavy-ion collisions, several algorithms do a good job in this respect \cite{Nonaka:2012qw,Molnar:2009tx}. These are, in particular,
the \textsc{shasta} algorithm \cite{SHASTA:1973}, which after finding a low-order solution with large numerical diffusion, corrects the result using
anti-diffusion fluxes in a second step, and the Kurganov-Tadmor (KT) algorithm \cite{Kurganov:2000}, an improvement of the Nessyahu-Tadmor algorithm (NT) 
\cite{Nessyahu:1990}, which is Riemann-solver free and can be viewed as an extension of the Lax-Friedrichs (LxF) scheme. Recently an algorithm with a relativistic Riemann solver geared towards simulating heavy-ion systems \cite{Nonaka:2012qw} has also been developed.
Comparisons of the performance of the first three types of algorithms in solving the relativistic Riemann problem were performed \cite{Molnar:2009tx} and
no clear difference was found in either performance or accuracy of the algorithms at a given lattice size. The analysis was extended \cite{Nonaka:2012qw} to include the Riemann solver based algorithm, which naturally is found to have some advantage in solving the Riemann problem in the test scenario.

\subsection{Recent results and comparison to experimental data}
Recently, event-by-event hydrodynamic calculations have produced a plethora of predictions and explanations of a wide range of 
experimental data, for which the inclusion of initial state fluctuations is the essential ingredient.
Not only do these simulations allow for the determination of all average higher flow harmonics beyond elliptic flow, but further details like
the event-by-event probability distributions of the flow harmonics or event plane correlations can be computed and compared to experimental data.

\begin{figure}[htb]
   \begin{center}
    \begin{minipage}{0.495\textwidth}
      \includegraphics[width=6.5cm]{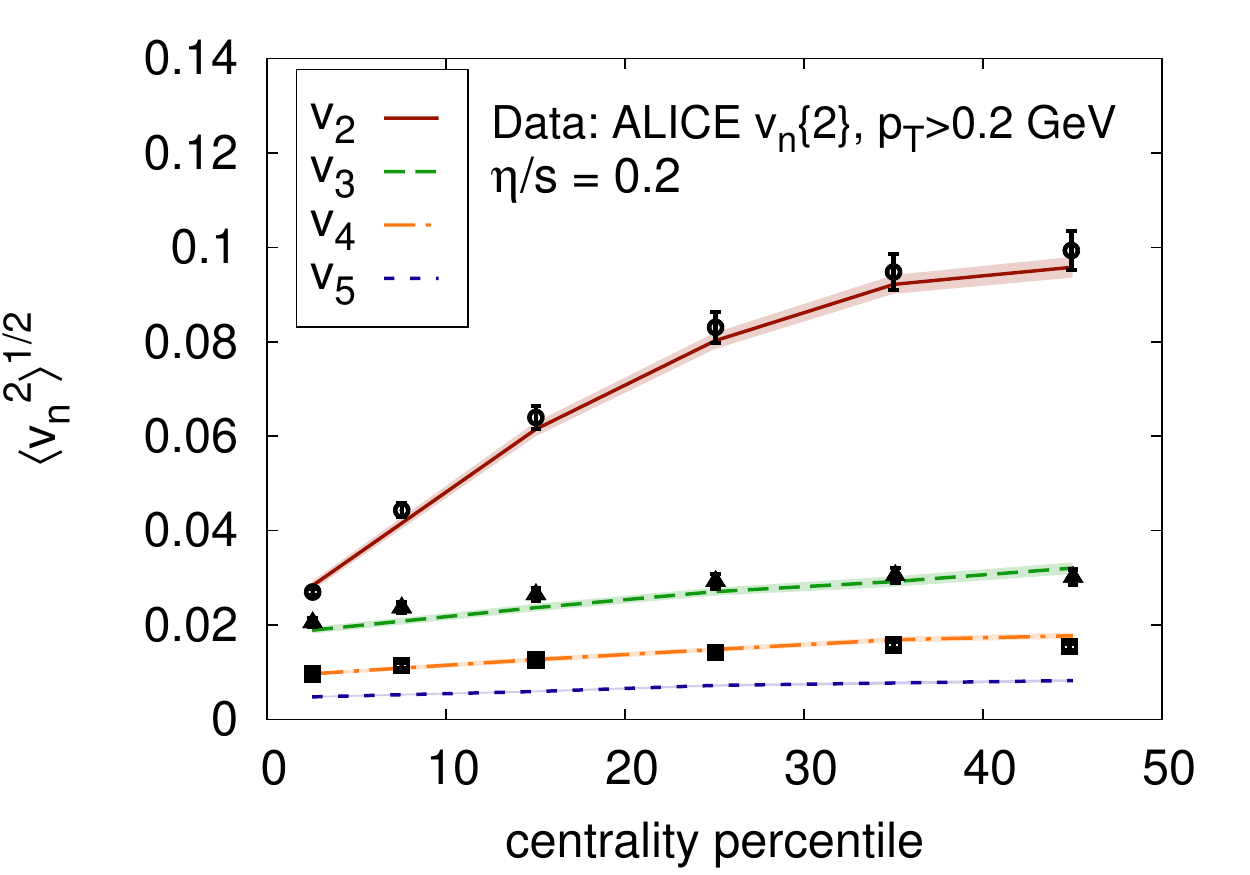} 
    \end{minipage}
    \hfill
    \begin{minipage}{0.495\textwidth}
      \vspace{0.1cm}
     \includegraphics[width=6.7cm]{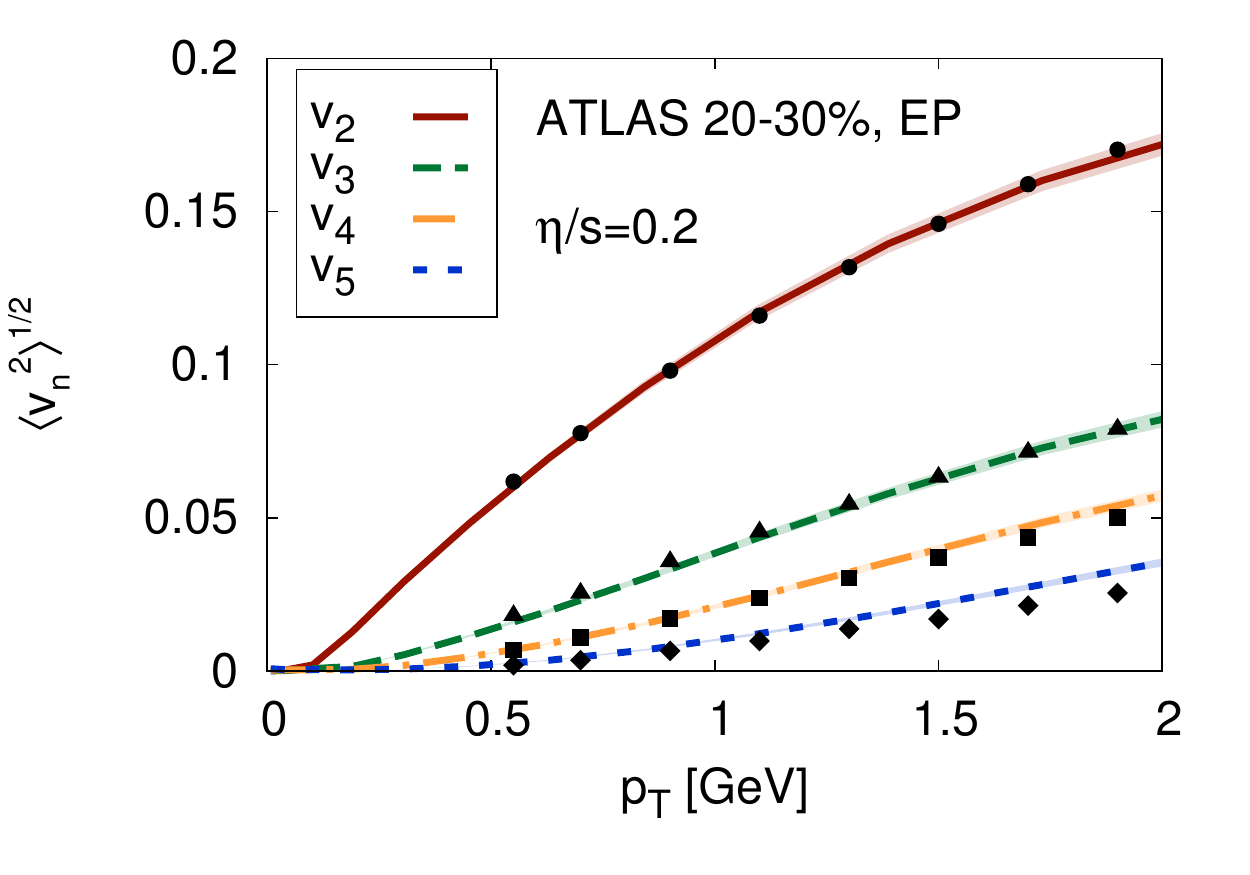}
    \end{minipage}
     \caption{Left: Root-mean-square anisotropic flow coefficients $\langle v_n^2 \rangle ^{1/2}$ in the IP-Glasma model \cite{Gale:2012rq}, computed as a function of centrality, 
       compared to experimental data of $v_n\{2\}$, $n\in\{2,3,4\}$, by the ALICE collaboration \cite{ALICE:2011ab} (points).
Right: Root-mean-square anisotropic flow coefficients $\langle v_n^2 \rangle ^{1/2}$ as a function of transverse momentum, 
       compared to experimental data by the ATLAS collaboration using the event plane (EP) method \cite{ATLAS:2012at} (points). Bands indicate statistical errors. 
     \label{fig:iglasma-vn}}
   \end{center}
\end{figure}

Simulations using the IP-Glasma initial state model have been particularly successful in describing both the $p_T$ dependent and integrated $v_n$ at both RHIC and LHC energies \cite{Gale:2012rq}. The agreement with experimental results from LHC shown in Fig. \ref{fig:iglasma-vn} is particularly striking.

This agreement indicates that initial state fluctuations in the deposited energy density, translated by hydrodynamic evolution into anisotropies 
in the particle production, are the main ingredient to explain the measured flow coefficients.

Because of this feature, some effort has been concentrated on characterizing the initial state in a way that ties it directly to the measured flow.
The simplest way of doing so is to compare the initial eccentricities of the system 
\begin{equation}
  \varepsilon_n = \frac{\sqrt{\langle r^n \cos(n\phi)\rangle^2+\langle r^n \sin(n\phi)\rangle^2}}{\langle r^n \rangle}
\end{equation}
to the final flow harmonics $v_n$. However, in particular for $v_4$ and higher harmonics, the nonlinear nature of hydrodynamics becomes important \cite{Gardim:2011xv} and more accurate predictors for flow coefficients involve both linear and nonlinear terms, e.g. $v_5$ has contributions from $\varepsilon_5$ and $\varepsilon_2 \varepsilon_3$, and it was shown \cite{Teaney:2012ke} that the nonlinear term becomes more dominant with increasing viscosity.

The fact that linear terms are damped more by viscosity leads to a growing correlation of different event planes 
\begin{equation}
  \psi_n=\frac{1}{n}\arctan\frac{\langle \sin(n\phi)\rangle}{\langle \cos(n\phi)\rangle}\,,
\end{equation}
with increasing viscosity \cite{Teaney:2012ke}, a result that is in line with findings in a different work \cite{Qiu:2012uy}, where experimental data on 
event plane correlations from the ATLAS collaboration \cite{Jia:2012sa} was compared to hydrodynamic calculations in different scenarios.

It has been argued that the most precise value for the shear viscosity to entropy density ratio $\eta/s$ of the quark-gluon plasma can be obtained 
by a fit to $p_T$-integrated $v_n$ measurements in ultra-central collisions, because in those fluctuations dominate all $v_n$ and uncertainties in 
$\varepsilon_2$, which varies most from one initial state model to the next, are minimal. \cite{Luzum:2012wu}
At the very least, the study of ultra-central collisions can constrain the possible fluctuations implemented in different models by comparison to experimental data.

\begin{figure}[htb]
   \begin{center}
    \begin{minipage}{0.45\textwidth}
     \includegraphics[width=6cm]{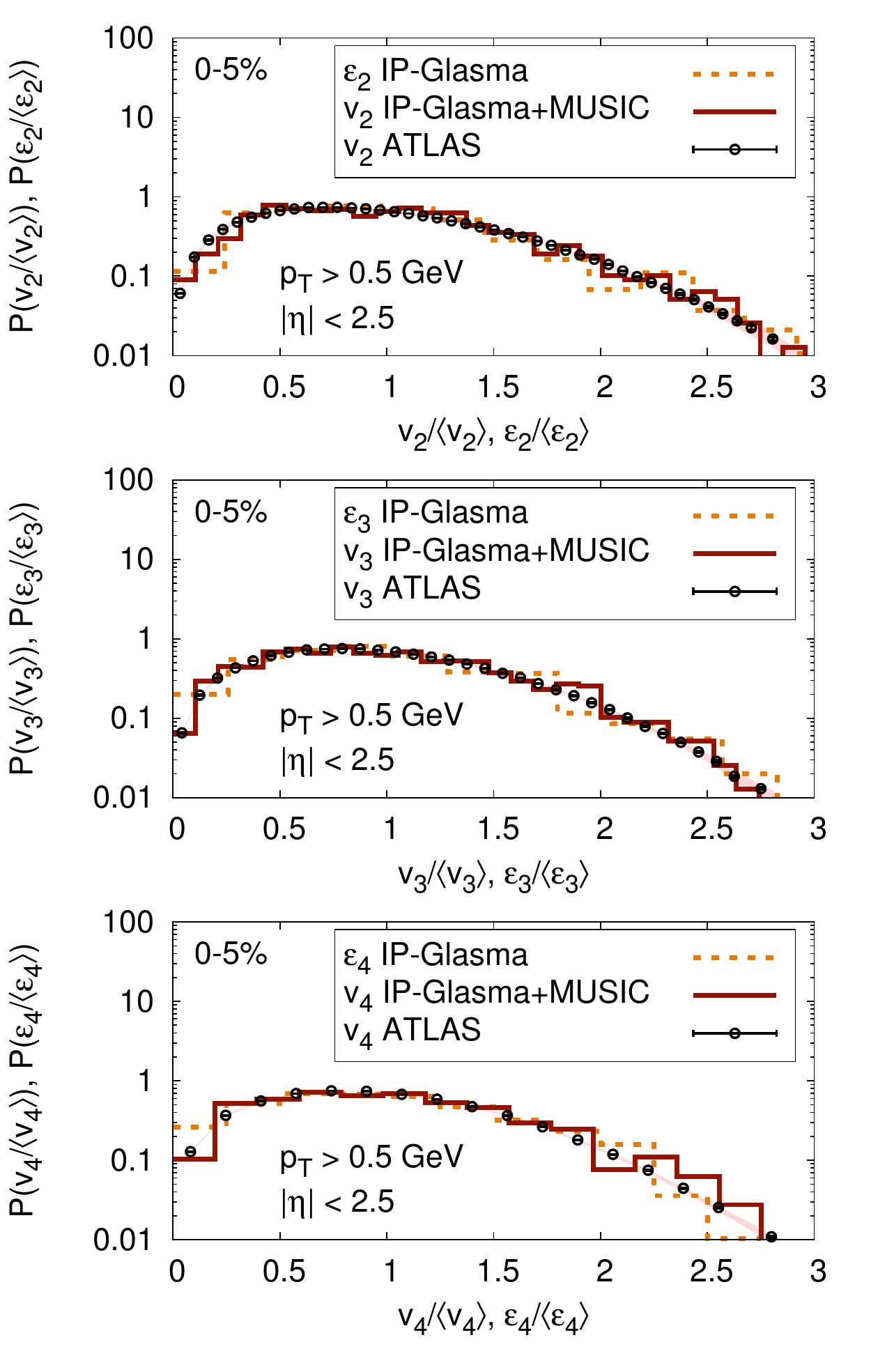}
    \end{minipage}
    \hfill
    \begin{minipage}{0.45\textwidth}
    \includegraphics[width=6cm]{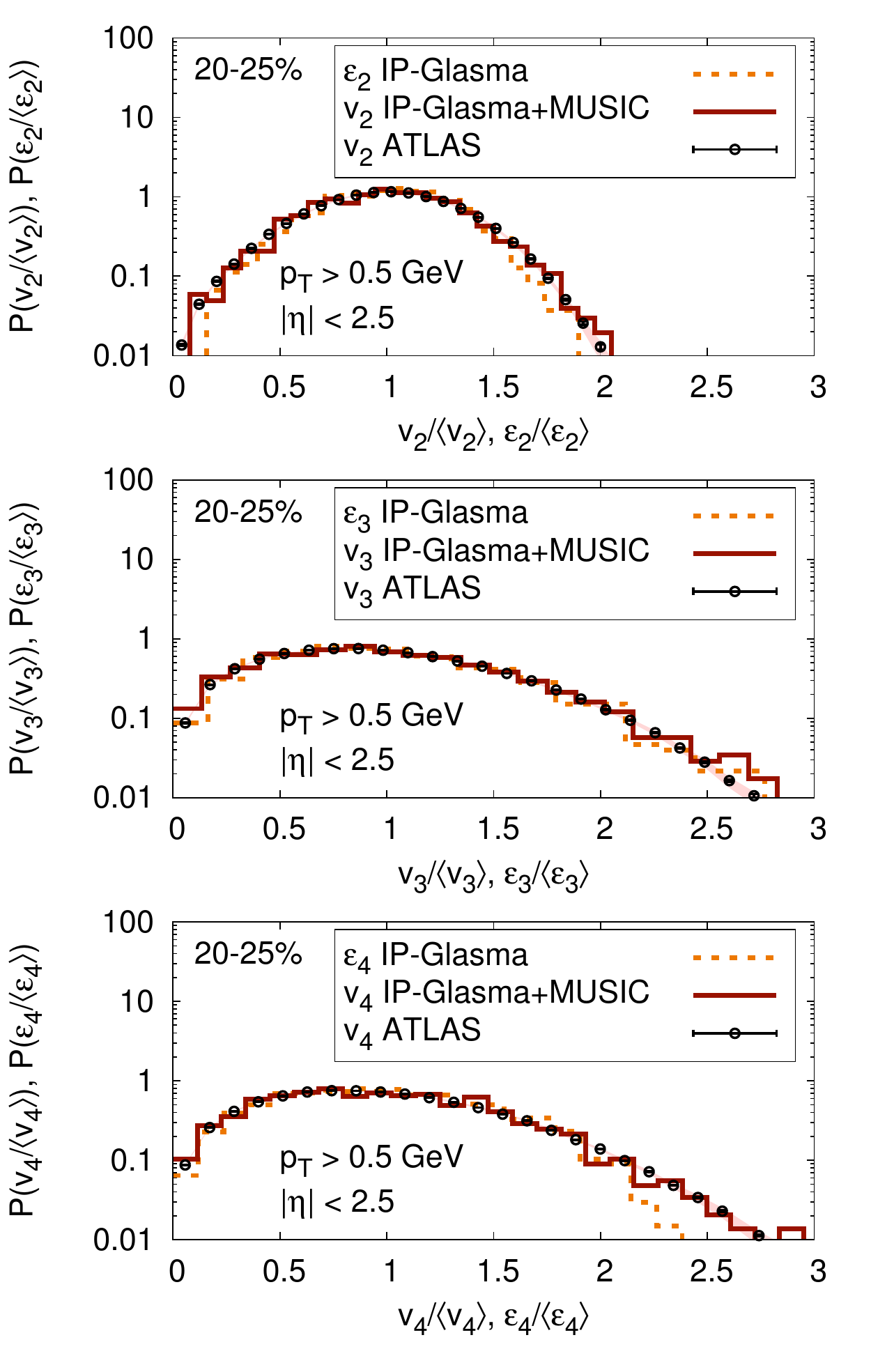}
    \end{minipage}
     \caption{Scaled distributions of $v_2$, $v_3$ and $v_4$ as well as $\varepsilon_2$, $\varepsilon_3$ and $\varepsilon_4$ from the IP-Glasma model \cite{Gale:2012rq} compared to experimental data from the ATLAS
     collaboration \cite{ATLAS:2012Jia,Jia:2012ve}. Using 750 (0-5\%) and 1300 (20-25\%) events. Bands are systematic experimental errors.}
     \label{fig:vnenDist-20-25}
   \end{center}
\end{figure}

Another way to constrain initial state models with fluctuations is the study of event-by-event distributions of flow harmonics $v_n$, which have recently be determined experimentally \cite{Jia:2012ve}. It was found \cite{Niemi:2012aj} that these distributions are almost independent of the details of the hydrodynamic evolution, like the shear viscosity to entropy density ratio. In fact, the distributions of initial eccentricities provide already an excellent approximation of the measured $v_n$ distributions when scaled by the mean value. This allows for an almost independent determination of the initial state model and the transport properties of the evolving system.
As shown in Fig. \ref{fig:vnenDist-20-25} the IP-Glasma model provides a very good description of these probability distributions \cite{Gale:2012rq} and differences between the $\varepsilon_n$ and $v_n$ distributions are only visible in the tail where nonlinear effects in the hydrodynamic evolution are potentially more important.

While most simulations take into account fluctuations of the energy density and flow velocities in the transverse plane only,
some studies have also considered fluctuations in the longitudinal direction \cite{Florchinger:2011qf,Pang:2012uw,Bzdak:2012tp},
finding a reduction of pion elliptic flow compared to calculations with only fluctuations in the transverse plane \cite{Pang:2012uw}.
This is an interesting possibility and needs to be investigated more closely in the future.

Event-by-event hydrodynamic calculations have also been utilized to compute dihadron correlations in heavy-ion collisions. When including local charge conservation, agreement with the two-dimensional two-particle correlation data in relative azimuthal angle and pseudorapidity at soft transverse momenta ($p_T < 2\,{\rm GeV}$) is found\cite{Bozek:2012en}. It should however be noted that the origin of the almost boost-invariant initial state, necessary to reproduce the measured ``ridge'' structure in pseudo-rapidity is not described by hydrodynamics - it is merely maintained by the hydrodynamic evolution, and requires an independent explanation.

Recent efforts have pushed these studies further to include $p+A$ and even $p+p$ collisions \cite{Bozek:2011if,Bozek:2012gr,Werner:2011fd}.
Whether the system size and viscous corrections in such small systems allow for the application of hydrodynamics is however questionable. 
Saturation models have reproduced correlation data over a wide kinematic range, \cite{Dusling:2012ig,Dusling:2012cg,Dusling:2012wy} showing that additional hydrodynamic flow is in fact required to describe the data from heavy-ion collisions, but has no room in smaller systems like $p+p$ \cite{Dusling:2012ig}.

\section{Summary, Conclusions and Outlook}
Relativistic viscous hydrodynamics has been extremely successful in describing the bulk properties of heavy-ion collisions at RHIC and LHC,
ranging from particle spectra to anisotropic flow and correlations. In particular the use of event-by-event hydrodynamic simulations together with models for the fluctuating initial state has dramatically increased the amount of successful predictions, such as the distributions of higher harmonics $v_n$, correlations of event planes, etc.
Comparison of experimental data to hydrodynamic simulations thus allows to extract properties of the matter created in heavy-ion collisions, such as the fact that the system is strongly interacting, and more quantitative measures like the shear viscosity to entropy density ratio $\eta/s$. Hydrodynamics is at this point in time the best tool for the determination of such fundamental properties of a hot and dense quantum-chromo dynamic system.

Despite these major successes, several unknowns and uncertainties remain. In particular it is not understood how the system reaches a thermalized state that is necessary for the usual hydrodynamic framework to be applicable. In fact, it is not settled whether the system actually does thermalize by the time we start applying hydrodynamics or whether momentum distributions remain largely anisotropic up to late times and hydrodynamics should be modified to accommodate this situation.

Related to this is the question of how to initialize viscous corrections and whether they become too large in fluctuating events with large gradients.
This would take us out of the domain of validity of hydrodynamics. The issue is particularly problematic when studying temperature dependent $\eta/s$ that can become very large at large ($T\gg T_c$) and small ($T<T_c$) temperatures. Temperature dependent values of additional transport coefficients such as bulk viscosity and relaxation times pose another large uncertainty that demands further detailed studies.

In addition there remain uncertainties in the freeze-out description, be it a simple Cooper-Frye description with resonance decays or coupling to a hadronic rescattering simulation. These lie for example in the determination and implementation of the freeze-out condition and the prescription of the conversion of energy densities to particle degrees of freedom.

Current research is addressing these and other caveats and will hopefully shed further light on the complex details of the
strongly interacting system created in heavy-ion collisions, and hence the properties of quantum-chromo dynamics under extreme conditions.


\section*{Acknowledgments}
We thank Raju Venugopalan for helpful comments on the manuscript and discussions.
This work was supported in part by the US Department of Energy under DOE Contract No.DE-AC02-98CH10886 
and in part by the Natural Sciences and Engineering Research Council of Canada.

\bibliography{spires}

\end{document}